\documentclass{aastex63}
\usepackage{graphicx}
\usepackage{gensymb}

\usepackage{graphicx}
\usepackage{wrapfig}
\usepackage{sidecap}
\usepackage{float}
\usepackage{array}
\usepackage{booktabs}
\usepackage{soul}
\usepackage{tablefootnote}
\shorttitle{A CO$_2$ cycle on Ariel?}
\shortauthors{Cartwright et al.}

\begin{document}
	
	\title{A CO$_2$ cycle on Ariel? \\
		Radiolytic production and migration to low latitude cold traps}

	\correspondingauthor{Richard J. Cartwright}
	\email{rcartwright@seti.org}
	
	\author[0000-0002-6886-6009]{Richard J. Cartwright$^a$}
	\affiliation{The Carl Sagan Center at the SETI Institute \\
		189 Bernardo Ave., Suite 200\\
		Mountain View, CA 94043, USA}
	\footnote{$^a$Visiting Astronomer at the Infrared Telescope Facility, which is operated by the University of Hawaii under contract 80HQTR19D0030 with the National Aeronautics and Space Administration. }
	
	\author{Tom A. Nordheim}
	\affiliation{Jet Propulsion Laboratory, California Institute of Technology \\
		4800 Oak Grove Drive\\
		Pasadena, CA 91109, USA}
	
	\author{David Decolibus}
	\affiliation{New Mexico State University \\
		PO Box 30001, MSC 4500\\
		Las Cruces, NM 88001, USA}
	
	\author{William  M. Grundy}
	\affiliation{Lowell Observatory \\
		1400 W Mars Hill Road\\
		Flagstaff, AZ 86001, USA}
	\affiliation{Northern Arizona University \\
		S San Francisco Street\\
		Flagstaff, AZ 86011, USA}
	
	\author{Bryan J. Holler}
	\affiliation{Space Telescope Science Institute \\
		3700 San Martin Drive\\
		Baltimore, MD 21218, USA}
	
	\author{Chloe B. Beddingfield}
	\affiliation{The Carl Sagan Center at the SETI Institute \\
		189 Bernardo Ave., Suite 200\\
		Mountain View, CA 94043, USA}
	\affiliation{NASA-Ames Research Center\\
		Mail Stop 245-1\\
		Building N245, Room 204\\
		P.O. Box 1\\
		Moffett Field, CA 94035, USA}
	
	\author{Michael M. Sori}
	\affiliation{Purdue University \\
		610 Purdue Hall \\
		West Lafayette, IN 47907, USA}
	
	\author{Michael P. Lucas}
	\affiliation{University of Tennessee\\
		1621 Cumberland Avenue\\
		Knoxville, TN 37996, USA}
	
	\author{Catherine M. Elder}
	\affiliation{Jet Propulsion Laboratory, California Institute of Technology \\
		4800 Oak Grove Drive\\
		Pasadena, CA 91109, USA}
	
	\author{Leonardo H. Regoli}
	\affiliation{John Hopkins University Applied Physics Laboratory \\
		11100 Johns Hopkins Road\\
		Laurel, MD 20723, USA}
	
	\author{Dale P. Cruikshank}
	\affiliation{NASA-Ames Research Center\\
		Mail Stop 245-1\\
		Building N245, Room 204\\
		P.O. Box 1\\
		Moffett Field, CA 94035, USA}
	
	\author{Joshua P. Emery}
	\affiliation{Northern Arizona University \\
		S San Francisco Street\\
		Flagstaff, AZ 86011, USA}
	
	\author{Erin J. Leonard}
	\affiliation{Jet Propulsion Laboratory, California Institute of Technology \\
		4800 Oak Grove Drive\\
		Pasadena, CA 91109, USA}
	
	\author{Corey J. Cochrane}
	\affiliation{Jet Propulsion Laboratory, California Institute of Technology \\
		4800 Oak Grove Drive\\
		Pasadena, CA 91109, USA}
	
	
	
	\begin{abstract}
		
		CO$_2$ ice is present on the trailing hemisphere of Ariel but is mostly absent from its leading hemisphere. The leading/trailing hemispherical asymmetry in the distribution of CO$_2$ ice is consistent with radiolytic production of CO$_2$, formed by charged particle bombardment of H$_2$O ice and carbonaceous material in Ariel's regolith. This longitudinal distribution of CO$_2$ on Ariel was previously characterized using 13 near-infrared reflectance spectra collected at `low' sub-observer latitudes between 30$\degree$S to 30$\degree$N. Here, we investigated the distribution of CO$_2$ ice on Ariel using 18 new spectra: two collected over low sub-observer latitudes, five collected at `mid' sub-observer latitudes  (31$\degree$ -- 44$\degree$N), and eleven collected over `high' sub-observer latitudes (45$\degree$ -- 51$\degree$N). Analysis of these data indicates that CO$_2$ ice is primarily concentrated on Ariel's trailing hemisphere. However, CO$_2$ ice band strengths are diminished in the spectra collected over mid and high sub-observer latitudes. This sub-observer latitudinal trend may result from radiolytic production of CO$_2$ molecules at high latitudes and subsequent migration of this constituent to low latitude cold traps. We detected a subtle feature near 2.13 $\micron$ in two spectra collected over high sub-observer latitudes, which might result from a `forbidden' transition mode of CO$_2$ ice that is substantially stronger in well mixed substrates composed of CO$_2$ and H$_2$O ice, consistent with regolith-mixed CO$_2$ ice grains formed by radiolysis. Additionally, we detected a 2.35-$\micron$ feature in some low sub-observer latitude spectra, which might result from CO formed as part of a CO$_2$ radiolytic production cycle.

	\end{abstract}
	
	\keywords{Uranian satellites (1750); Planetary surfaces (2113); 
		Surface composition (2115); Surface processes (2116); Surface ices (2117)}
	
	
	\section{Introduction} 
	Ariel and the other `classical' Uranian moons Miranda, Umbriel, Titania, and Oberon are candidate ocean worlds \citep[e.g.,][]{hendrix2019nasa,cartwright2021sciencecase} with surfaces that have been modified by endogenic geologic processes \citep[e.g.,][]{beddingfield2015fault,beddingfield2020hidden,beddingfield2021Arielcryo,schenk2020topography} and interactions with the surrounding space environment \citep[e.g.,][]{grundy2006distributions,cartwright2018red,decolibus2020investigating}. Ground-based, near-infrared (NIR) observations have determined that the surfaces of these moons are primarily composed of H$_2$O ice mixed with dark and spectrally neutral constituents \citep[e.g.,][]{cruikshank1977identification,cruikshank1980near,cruikshank1981uranian,soifer1981near,brown1983uranian,brown1984surface} that likely include a carbon-rich component \citep[]{clark1984spectral}. CO$_2$ ice has also been detected
	on Ariel, Umbriel, Titania, and Oberon via three prominent absorption features
	centered near 1.966, 2.012, and 2.070 $\micron$ \citep[]{grundy2003discovery,grundy2006distributions,cartwright2015distribution}. These bands are stronger on the trailing hemispheres of these moons (longitudes 181 -- 360$\degree$) compared to their leading hemispheres (longitudes 1 -- 180$\degree$). Furthermore, these CO$_2$ ice bands are strongest on Ariel and get progressively weaker with increasing orbital radius, with the weakest CO$_2$ bands detected on the outermost classical moon Oberon \citep[e.g.,][]{cartwright2015distribution}. 
	
	Previous studies have suggested that the observed longitudinal and radial trends in the distribution of CO$_2$ ice are consistent with a radiolytic origin \citep[]{grundy2006distributions,cartwright2015distribution}. In this scenario, charged particles from Uranus' magnetosphere bombard native H$_2$O ice and carbonaceous material exposed on the surfaces of the classical moons, driving radiolytic production of CO$_2$ molecules. Because Uranus rotates faster ($\sim$17.2 hours) than the orbital periods of its classical moons ($\sim$1.4 -- 13.5 days), charged particles trapped in its magnetosphere should preferentially interact with their trailing hemispheres as Uranus' magnetic field lines sweep past these moons. Voyager 2 observed depletions in magnetospheric charged particle flux near the orbital location of some of the Uranian moons (including Ariel), hinting at extensive interactions between Uranus' magnetosphere and the surfaces of its moons, possibly driving radiolytic chemistry \citep[e.g.,][]{paranicas1996charged}. Alternatively, CO$_2$ ice could be a native constituent on the classical moons that is sourced from their interiors by geologic processes that expose and/or emplace material rich in CO$_2$, perhaps similar to Enceladus where plume activity has likely deposited large amounts of CO$_2$ on its surface \citep[e.g.,][]{waite2006cassini,combe2019nature}.
	
	The spectral signatures of the detected CO$_2$ bands are remarkably consistent with `pure' CO$_2$ ice measured in the laboratory (i.e., segregated from other constituents in concentrated deposits with crystal structures dominated by CO$_2$ molecules) \citep[e.g.,][]{hansen1997spectral,gerakines2005strengths}. At the estimated peak surface temperatures of the Uranian moons (80 -- 90 K, \citealt[]{hanel1986infrared,sori2017wunda}), CO$_2$ ice is thermodynamically unstable over geologic timescales \citep[]{grundy2006distributions,sori2017wunda}. Because of Uranus' large obliquity ($\sim$98$\degree$), the poles of the classical moons are bathed in continuous sunlight during the Uranian system's long summers. As a result, long-term cold traps for CO$_2$ ice are likely concentrated at low latitudes, where the timescales of maximum solar heating are much shorter compared to high latitudes \citep[]{grundy2006distributions,sori2017wunda}. Additionally, CO$_2$ molecules should concentrate at similar longitudes to where they were formed or exposed \citep[]{sori2017wunda}, likely explaining why CO$_2$ ice remains concentrated on the trailing hemisphere of Ariel and the other Uranian moons. Thermodynamical models therefore predict that CO$_2$ molecules generated or exposed at polar latitudes should sublimate, migrate to equatorial latitudes, and condense in cold traps, primarily on Ariel's trailing hemisphere. 
	
	Previously analyzed NIR spectra were collected during late southern summer and northern spring (subsolar latitudes 30$\degree$S -- 25$\degree$N), when low latitude regions that are likely rich in CO$_2$ (30$\degree$S -- 30$\degree$N) represented $>$ 50$\%$ of these moons' observed disks. Consequently, modeling results that suggest CO$_2$ could be generated/exposed at high latitudes and then migrate to low latitude cold traps have not been tested using spectra collected at higher sub-observer latitudes that better sample polar regions. We present new results on the distribution of CO$_2$ ice on Ariel, the Uranian moon with the strongest CO$_2$ ice bands, using NIR spectra collected over `mid' sub-observer latitudes (31 -- 44$\degree$N) and `high' sub-observer latitudes (45$\degree$ -- 51$\degree$N) latitudes, thereby expanding on the latitudinal coverage provided by previously collected spectra.
	
	\section{Data and Methods}
	
	\subsection{Observations and Data Reduction} 
	We analyzed 31 NIR reflectance spectra, ten of which are reported here for the first time, with the other 21 spectra reported previously (observation details summarized in Table 1) \citep[]{grundy2003discovery,grundy2006distributions,cartwright2015distribution,cartwright2018red,cartwright2020evidence}. Four of the new spectra, and all 21 previously reported spectra, were collected using the NIR SpeX spectrograph/imager at NASA's Infrared Telescope Facility (IRTF), operating in moderate resolution short cross-dispersed (SXD) mode ($\sim$0.8 -- 2.5 $\micron$) \citep[e.g.,][]{rayner2003spex}. The other six new spectra were collected using the TripleSpec spectrograph (0.95 -- 2.46 $\micron$) on the Astrophysical Research Consortium (ARC) 3.5-m telescope at the Apache Point Observatory \citep[]{wilson2004mass}. 
	
	When possible, observations were made using slit orientations that matched or were similar to the parallactic angle to minimize atmospheric dispersion, which can introduce spectral slope changes to the continua of spectra over visible and near-infrared wavelengths, in particular at wavelengths $\lesssim$ 1.2 $\micron$. Some of the observations of Ariel were made using slit orientations that deviated notably from the parallactic angle to minimize scattered light from the bright disk of Uranus. Nevertheless, it is likely that atmospheric dispersion over the wavelength range of the CO$_2$ ice bands between 1.5 to 1.7 $\micron$ did not exceed 0.05 and was typically $<$ 0.02  (relative to 2.2 $\micron$). Atmospheric dispersion over the wavelength range of the CO$_2$ ice bands between 1.9 to 2.1 $\micron$ did not exceed 0.02 and was typically $<$ 0.01  (relative to 2.2 $\micron$). Consequently, the impact of atmospheric dispersion on our measurements and modeling of the spectral continuum proximal to these CO$_2$ ice bands is likely negligible.
	
	IRTF/SpeX observations were made by placing Ariel in two different positions (`A' and `B') separated by 7.5'' on a 15'' long slit. The resulting exposures were separated into sequential `AB' pairs. We then subtracted the B  exposures from the A exposures to provide a first order correction for sky emission. The resulting A-B subtracted pairs were flatfielded to account for variations across the detector. Flat frames were generated by illuminating SpeX's internal integrating sphere with a quartz lamp. Wavelength calibration was performed using argon emission lines. We used the Spextool data reduction suite \citep{cushing2004spextool}, along with custom programs, to calibrate and extract all Ariel spectra. To remove the solar spectrum, perform additional atmospheric correction, and remove instrument artifacts, all Ariel spectra were divided by solar analog star spectra, which were observed on the same night and close in time and airmass to the observations of Ariel  (within $\pm 0.1$ airmass). We observed HD 12124 (G0) and HD 16275 (G5) in 2020. To increase signal-to-noise (S/N), all star-divided frames from the same night were co-added, and the resulting uncertainties were calculated using the standard error ($\sigma$/$\sqrt{n}$) of each co-added pixel. The other 21 SpeX spectra we analyzed were collected by different teams between 2000 and 2019  \citep{grundy2003discovery, grundy2006distributions, cartwright2015distribution, cartwright2018red,cartwright2020evidence}.
	
	Spectra collected with ARC 3.5m/TripleSpec were obtained in a similar manner. The A and B positions are separated by 21'' along a 1.1x43'' slit. The length of the TripleSpec slit sometimes allowed for opportunities to place multiple Uranian moons in the slit simultaneously. In these cases, the moons did not fall exactly on the nominal A and B positions, but the spectra were reduced and extracted using the same A - B procedure. Data reduction was performed using TriplespecTool, a modified version of Spextool \citep{cushing2004spextool}. Flat frames were obtained with quartz lamps mounted on the telescope structure. We subtracted equivalent length, unilluminated exposures from the lamp flats to remove intrinsic telescope thermal emission. Wavelength calibration was performed using OH airglow emission lines in our science frames. For our telluric correction, we observed the solar analog stars HD 16017 (G2V), BD+15 4915 (G2V), HD 19061 (G2V), and HD 224251 (G2V), in 2019 and 2020, and HD 16275 (G5) in 2021, usually within $\pm 0.1$ airmass of our observations of Ariel.

	\begin{table}[]
		\caption {IRTF/SpeX and ARC 3.5-m/TripleSpec observations of Ariel.} 
		\hskip-0.8cm\begin{tabular}{*9c}
			\hline\hline
			\begin{tabular}[c]{@{}l@{}}\hspace{-1 cm}Sub-\\  \hspace{-1 cm}Observer\\  \hspace{-1 cm}Long. ($\degree$)\end{tabular} & \begin{tabular}[c]{@{}l@{}} \hspace{-1 cm}Sub-\\  \hspace{-1 cm}Observer \\  \hspace{-1 cm}Lat.  ($\degree$)\end{tabular} & UT Date & \begin{tabular}[c]{@{}l@{}} \hspace{-1 cm}UT Time  \\  \hspace{-1 cm}(mid-expos)\end{tabular} & \begin{tabular}[c]{@{}l@{}} \hspace{-1 cm}Integration \\ \hspace{-1 cm}Time   (min)\end{tabular} & \begin{tabular}[c]{@{}l@{}} \hspace{-1 cm}Spectrograph \end{tabular} & \begin{tabular}[c]{@{}l@{}} \hspace{-1 cm}Slit \\  \hspace{-1 cm}Width \\  \hspace{-1 cm}('')\end{tabular} & \begin{tabular}[c]{@{}l@{}} \hspace{-1 cm}Average \\  \hspace{-1 cm}Resolving \\  \hspace{-1 cm}Power ($\lambda$/$\Delta$$\lambda$)\end{tabular} & References \\
			\hline
			6.8 & 44.8 & 1/18/20& 4:20 & 76 & TripleSpec & 1.1 & 3500 & This work \\
			15.3 & 27.8 & 9/15/14 & 11:35 & 88 & SpeX & 0.8& 750 & Cartwright et al. (2018) \\
			26.7 & 48.8& 1/5/21 & 3:55 & 60 & TripleSpec & 1.1 & 3500 & This work \\
			53.6 & -16.0 & 8/9/03 & 12:15 & 156 & SpeX & 0.3& 2000 & Grundy et al. (2006) \\
			79.8 & -19.4 & 7/17/02 & 13:25 & 108 & SpeX & 0.5& 1200 & Grundy et al. (2003) \\
			87.8 & 24.0 & 9/5/13 & 11:10 & 92 & SpeX & 0.8& 750 & Cartwright et al. (2015) \\
			93.5 & -18.1 & 10/4/03 & 5:45 & 108 & SpeX & 0.3& 2000 & Grundy et al. (2006) \\
			100.3 & 46.4& 11/4/19 & 5:10& 96 & TripleSpec & 1.1 & 3500 & This work \\		
			110.1 & 32.0 & 9/11/15 & 13:30 & 44 & SpeX & 0.8& 750 & Cartwright et al. (2018) \\
			132.2 & 28.5 & 8/24/14 & 14:05 & 40 & SpeX& 0.8& 750 & Cartwright et al. (2018) \\
			144.8 & 43.2 & 10/12/18 & 9:30 & 73 & SpeX & 0.5& 1200 & Cartwright et al. (2020b)  \\
			152.9 & 51.2 & 10/17/20 & 9:20 & 60 & SpeX & 0.5& 1200 & This work \\
			159.9 & -11.1 & 7/15/04 & 12:00 & 112 & SpeX & 0.3& 2000 & Grundy et al. (2006) \\
			200.0 & -15.9 & 8/5/03 & 12:00 & 84 & SpeX & 0.3& 2000 & Grundy et al. (2006) \\
			205.2 & 50.5 & 11/4/20 & 9:30 & 84 & SpeX & 0.8 & 750 & This work \\		
			205.5 & 46.3 & 11/7/19 & 11:20 & 60 & SpeX & 0.5& 1200 & Cartwright et al. (2020b) \\
			219.8 & -17.2 & 9/7/03 & 9:35 & 90 & SpeX & 0.3& 2000 & Grundy et al. (2006) \\
			231.8 & 51.1 & 10/20/20 & 11:00 & 60 & SpeX & 0.8& 750 & This work \\
			233.8 & -23.1 & 7/5/01 & 14:10 & 50 & SpeX & 0.5& 1200 & Grundy et al. (2003) \\
			235.9 & 47.3 & 10/13/19 & 11:30 & 32 & TripleSpec & 1.1 & 3500 & This work \\
			257.6 & -29.5 & 9/6/00 & 7:35 & 76 & SpeX & 0.8& 750 & Cartwright et al. (2015) \\
			260.2 & 47.0 & 10/21/19 & 5:05 & 36 & TripleSpec & 1.1 & 3500 & This work \\	
			263.7 & 46.8 & 10/26/19 & 6:40 & 32 & TripleSpec & 1.1 & 3500 & This work \\	
			268.3 & 39.0 & 10/15/17 & 8:00 & 40 & SpeX & 0.5& 1200 & Cartwright et al. (2020b) \\
			273.2 & 42.2 & 11/7/18 & 12:00 & 42 & SpeX & 0.5& 1200 & Cartwright et al. (2020b) \\
			278.3 & 24.8 & 8/7/13 & 13:20 & 44 & SpeX & 0.8& 750 & Cartwright et al. (2015) \\
			292.8 & 51.2 & 10/18/20 & 8:50 & 84 & SpeX & 0.5 & 1200 & This work \\			
			294.8 & 39.5 & 9/30/17 & 9:30 & 120 & SpeX & 0.5& 1200 & Cartwright et al. (2020b) \\
			294.8 & -19.3 & 7/16/02 & 13:10 & 140 & SpeX & 0.5& 1200 & Grundy et al. (2003) \\
			304.8 & -23.2 & 7/8/01 & 14:40 & 48 & SpeX & 0.5& 1200 & Grundy et al. (2003) \\
			316.6 & -18.2 & 10/8/03 & 7:55 & 132 & SpeX & 0.3& 2000 & Grundy et al. (2006) \\
			\hline
		\end{tabular} 
	\end{table}
	
	\subsection{Band Parameter Analyses} 
	We analyzed three combination and overtone CO$_2$ ice absorption features centered near 1.966, 2.012, and 2.070 $\micron$, which we hereafter refer to as CO$_2$ bands 1, 2, and 3, respectively. To assess the distribution of CO$_2$ ice on Ariel, we measured the areas of these three CO$_2$ bands in each spectrum, using a custom program \citep{cartwright2015distribution}. The program identified the continua for each CO$_2$ band (example shown in Figure A1), divided each CO$_2$ band by its continuum, and used the trapezoidal rule to measure the areas of the resulting continuum-divided bands. To estimate band area errors, the program used Monte Carlo simulations to resample the 1$\sigma$ uncertainties for the spectral channels covered by each CO$_2$ band (iterated 20,000 times). After measuring all three CO$_2$ bands in a spectrum, the program summed them into one total band area and propagated errors.

\subsection{Radiative transfer modeling}
	To complement our CO$_2$ ice band area measurements, and provide additional context on the spectral signature of CO$_2$ ice on Ariel, we utilized a Hapke-Mie spectral modeling program to generate synthetic spectra of the grand average low, mid, and high sub-observer latitude spectra collected over Ariel's trailing hemisphere (shown in Figure 1). These synthetic spectra are one-layer models that provide an estimate of the fractional area of Ariel's trailing hemisphere covered by CO$_2$ ice, as well as providing additional context on the spectral signature of this constituent. In order to focus our modeling efforts on the spectral signature of CO$_2$ ice, we generated synthetic spectra spanning 1.5 to 1.7 $\micron$ and 1.9 to 2.1 $\micron$. 
	
	Mie theory, which is used to model scattering and absorption by randomly spaced spherical particles, can simulate scattering off particles that are smaller than or similar in size to the incident wavelength of light \citep[e.g.,][]{bohren1983light}. Mie scattering theory is useful for simulating planetary regoliths that include small grains like the Uranian moons' regoliths, which could include a large number of sub-micron to micron sized particles \citep[e.g.,][]{afanasiev2014polarimetry,cartwright2020probing}. In contrast, pure Hapke approaches provide less robust models for planetary regoliths that include grains comparable to, or smaller than, the wavelength of incident light \citep[e.g.,][]{emery2006thermal}. Our hybrid Hapke-Mie technique takes the complex indices of refraction (i.e., optical constants) for each constituent and calculates the single scattering albedo ($\bar{\omega}$$_0$) using Mie theory. The $\bar{\omega}$$_0$ values, along with the bidirectional reflectance, opposition effect, and the phase function, are utilized to calculate the geometric albedo of the synthetic spectrum \citep[e.g.,][]{hapke2012theory}. Small resonances can occur in scattering models that utilize Mie theory. To account for these resonances, our spectral modeling software incorporates a range of grain diameters (typically 10$\%$ spread in sizes) that are averaged together to match the desired grain size of each constituent in the simulated planetary regolith. 
	
	This spectral modeling program was used previously to generate best fit synthetic spectra for the CO$_2$ ice bands detected in Uranian moon reflectance spectra collected at low sub-observer latitudes \citep[]{cartwright2015distribution}. These synthetic spectra were generated using optical constants for crystalline H$_2$O ice (80 K) \citep{mastrapa2008optical}, crystalline CO$_2$ ice ($\sim$150 K) \citep{hansen1997spectral}, and amorphous carbon \citep{rouleau1991shape}, which simulates the dark, spectrally-neutral absorber that is present on the Uranian moons \citep[e.g.,][]{clark1984spectral}. Using the same approach as these previous modeling efforts, we generated particulate mixtures of  H$_2$O ice and amorphous carbon grains to simulate a well mixed regolith dominated by `dirty' H$_2$O ice where CO$_2$ molecules could be formed via radiolysis. We then linearly combined these dirty H$_2$O ice models with CO$_2$ ice to simulate concentrated deposits of pure CO$_2$ ice where migrating CO$_2$ molecules might get cold trapped. 
	
	Prior spectral modeling efforts demonstrate that best fit synthetic spectra for Ariel and the other Uranian moons include H$_2$O ice grains with three different grain diameters: a small fraction ($<$ 1$\%$) of sub-micron sized grains (typically 0.2, 0.3, or 0.5 $\micron$ diameters) and larger fractions of two different moderately sized H$_2$O ice grains, typically with 10 and 50 $\micron$ grain diameters \citep{cartwright2015distribution,cartwright2018red,cartwright2020probing}. The inclusion of multiple H$_2$O ice grain sizes is required to simultaneously fit the 1.52-$\micron$ and 2.02-$\micron$ H$_2$O ice bands, in particular small amounts of sub-micron H$_2$O ice grains, which can significantly alter the relative strengths of these two H$_2$O ice bands in the resulting synthetic spectrum. The synthetic spectra typically only require one amorphous C grain diameter (ranging between 5 and 12.5 $\micron$) to simulate the low albedo absorber that helps obscure shorter wavelength H$_2$O bands and `flattens' the spectral continuum between 1.4 and 2.4 $\micron$. Of note, prior work has demonstrated that amorphous pryoxene can also provide a suitable proxy for the dark, spectrally neutral absorber on the surfaces of the Uranian moons \citep{cartwright2018red}. Nevertheless, we included amorphous C in the synthetic spectra present here to better simulate a regolith where CO$_2$ ice is generated by radiolysis of native H$_2$O ice and carbonaceous species. Prior spectral modeling work demonstrated that a blend of two different CO$_2$ ice grain diameters provides a good match to the band shape of the detected CO$_2$ ice bands, with one smaller grain size (1 or 10 $\micron$ diameters) and another larger grain size (50, 100, or 200 $\micron$ diameters). Additional information on how grain size can modify synthetic spectra using this Hapke-Mie modeling program was reported in Appendix B of \citet{cartwright2015distribution}. Because the CO$_2$ ice bands between 1.5 and 1.7 $\micron$ require thick substrates to detect and measure in the laboratory ($\sim$2 -- 107 mm, \citealt{hansen2005ultraviolet}), and because CO$_2$ ice on Ariel could be concentrated in large low latitude cold traps, it is assumed to be optically thick (i.e., obscuring the spectral signature of Ariel's regolith beneath it). 

The quality of the fits provided by the synthetic spectra was assessed using reduced Chi Square statistics: $\chi_R^2$ = $\frac{1}{a}$ $\sum$ ($O_i$ - $M_i$)$^2$/$\sigma_i^2$, where \textit{a} is the degrees of freedom, \textit{$O_i$} is the observed data, \textit{$M_i$} is the modeled data, and $\sigma_i^2$ is the variance \citep[e.g.,][]{bevington1969data}. Typically, $\chi_R^2$ $\approx$ 1 indicates that a given model is a good match to a set of observed data, whereas $\chi_R^2$ $\gg$ 1 indicates that a given model is likely a poor match to a set of observed data. A $\chi_R^2$ $\ll$ 1 can also indicate a good fit to the data and may result from overestimated measurement errors. Additionally, fitting models to data with large uncertainties, as well as uncertainty in the degrees of freedom \citep{andrae2010and}, can lower the accuracy of $\chi_R^2$ statistics. 

All of the synthetic spectra reported here provide non-unique solutions for estimating the fractional coverage and grain size of the different constituents on Ariel's surface. Small adjustments in the grain sizes of different constituents included in each synthetic spectrum can be accommodated with complementary adjustments in the fractional coverage of different constituents (i.e., there is a degeneracy between grain size and fractional coverage). However, as demonstrated by prior work \citep{cartwright2015distribution,cartwright2018red,cartwright2020probing}, synthetic spectra that deviate substantially from the mixing regimes, constituents, and grain sizes described above do not provide good fits to the collected spectra of the Uranian moons. 
	
	\section{Results and Analyses} 
	
	\subsection{Near-infrared spectra of Ariel} 
	Ten new disk-integrated reflectance spectra of Ariel are shown in Figure A2. Nine of these spectra display evidence for CO$_2$ bands 1, 2, and 3, including all seven spectra collected over Ariel's trailing hemisphere (examples of CO$_2$ bands 1, 2, and 3 are shown in Figure 1). We also observe the weak CO$_2$ ice combination and overtone features between 1.5 and 1.7 $\micron$, which were first reported in \citealt{grundy2006distributions}, in some of the spectra collected over Ariel's trailing hemisphere. Additionally, we observe a subtle band near 2.13 $\micron$ in two of the new spectra, collected at similar sub-observer latitudes (47.3$\degree$N and 51.1$\degree$N) and longitudes (231.8$\degree$ and 235.9$\degree$) (Figure 4). The detected 2.13-$\micron$ band may result from a `forbidden' overtone mode of CO$_2$ ice, which is very weak in pure CO$_2$ ice deposits and much stronger in substrates composed of CO$_2$ intimately mixed with H$_2$O ice or methanol \citep[]{bernstein2005near}. We also observe a subtle band centered near 2.35 $\micron$ in some of the spectra collected over Ariel's trailing hemisphere (Figure 5). This feature might result from CO generated as part of a radiolytic production cycle of CO$_2$ ice \citep[e.g.,][]{gerakines2001energetic,zheng2007formation}. We discuss the 2.13-$\micron$ and 2.35-$\micron$ absorption features in greater detail in Section 4.1.
	
	\begin{figure}[h]
		\includegraphics[scale=0.92]{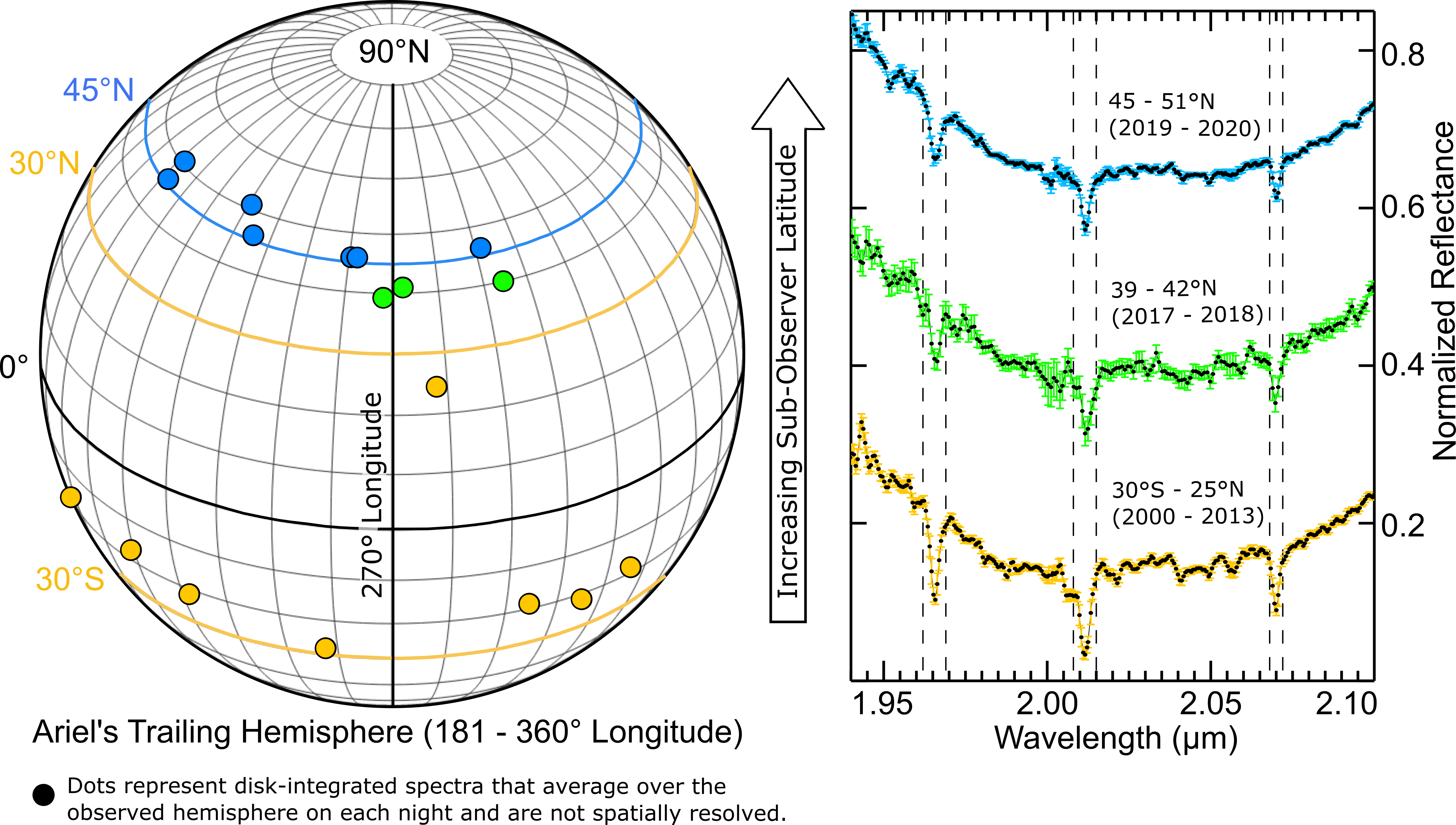}
		\caption{\textit{Left: Diagram showing the mid-observation, sub-observer longitudes and latitudes of reflectance spectra collected in the low (30$\degree$S -- 30$\degree$N), mid (31$\degree$ -- 44$\degree$N), and high (45$\degree$ -- 51$\degree$N) sub-observer latitude zones over Ariel's trailing hemisphere (gold, green, and blue dots, respectively). Ariel's Equator (0$\degree$) and the center of its trailing hemisphere (270$\degree$ longitude) are indicated (black lines). All collected spectra are disk-integrated and average over an entire hemisphere. Right: `Grand average' spectra and 1$\sigma$ uncertainties for Ariel's low, mid, and high sub-observer latitude zones (gold, green, and blue, error bars respectively). The spectra were normalized to 1 using the mean reflectance between 1.74 and 1.77 $\micron$ and then offset vertically for clarity. The sub-observer latitude range for the individual spectra, and the years they were collected in, are listed above each of the grand average spectra (see Table 1 for observation details). The wavelength ranges of CO$_2$ bands 1,2, and 3 are indicated (black dashed lines). The spectra were lightly smoothed using a 2 to 4 pixel wide boxcar function.}}\vspace{0.1 cm}
	\end{figure} 
	
	All ten new spectra display the 1.52-$\micron$ and 2.02-$\micron$ H$_2$O ice bands and the 1.65-$\micron$ crystalline H$_2$O ice feature detected previously on Ariel \citep[e.g.,][]{grundy2006distributions,cartwright2018red}. We also observe an absorption band centered near 2.2-$\micron$ in several of the new Ariel spectra, which has been detected previously and attributed to NH$_3$ and NH$_4$-bearing species \citep[]{cartwright2018red,cartwright2020evidence,cook2018composition}. Analysis of these H$_2$O ice bands and possible NH$_3$ and NH$_4$-bearing constituents is beyond the scope of this paper and will be included in future work.
	
	\subsection{CO$_2$ ice band parameter analyses} 
	We measured the areas of CO$_2$ bands 1, 2, and 3  in the ten new Ariel spectra, as well as eight previously reported Ariel spectra \citep[]{cartwright2018red,cartwright2020evidence} that do not have published CO$_2$ band area measurements. We report these 18 sets of CO$_2$ band areas and their 1$\sigma$ uncertainties in Table 2, along with CO$_2$ ice band areas for another 13 Ariel spectra that were measured previously \citep[]{grundy2003discovery,grundy2006distributions,cartwright2015distribution}. All 31 of the Ariel spectra were analyzed using the same CO$_2$ ice band measurement procedure, originally presented in \citet{cartwright2015distribution}. To investigate possible latitudinal trends in the distribution of CO$_2$ ice, we calculated mean CO$_2$ ice band areas using spectra collect at low (30$\degree$S -- 30$\degree$N),  mid (31 -- 44$\degree$N), and
	high (45$\degree$ -- 51$\degree$N) sub-observer latitudes (final column in Table 2). 
	
	\begin{table}[]
		\caption {CO$_2$ ice band areas.} 
		\hskip-0.9cm\begin{tabular}{*6c|c}
			\hline\hline
			\begin{tabular}[c]{@{}l@{}}\hspace{-1 cm}Sub-Observer \\ \hspace{-1 cm}Long. ($\degree$)\end{tabular} & \begin{tabular}[c]{@{}l@{}} \hspace{-1 cm}Sub-Observer \\  \hspace{-1 cm}Lat.  ($\degree$)\end{tabular}  & \begin{tabular}[c]{@{}l@{}} \hspace{-1 cm}CO$_2$ Band 1 \\  \hspace{-1 cm} (10$^-$$^4$$\micron$) \end{tabular} & \begin{tabular}[c]{@{}l@{}} \hspace{-1 cm}CO$_2$ Band 2 \\  \hspace{-1 cm} (10$^-$$^4$$\micron$) \end{tabular} & \begin{tabular}[c]{@{}l@{}} \hspace{-1 cm}CO$_2$ Band 3 \\  \hspace{-1 cm} (10$^-$$^4$$\micron$) \end{tabular} & \begin{tabular}[c]{@{}l@{}} \hspace{-1 cm}$^\dagger$CO$_2$ Total \\  \hspace{-1 cm}Area (10$^-$$^4$$\micron$) \end{tabular} & \begin{tabular}[c]{@{}l@{}} \hspace{-1 cm}$^\dagger$$^\dagger$Mean CO$_2$ Total \\  \hspace{-1 cm}Areas (10$^-$$^4$$\micron$) \end{tabular} \\
			\hline
			\textit{Leading Hemisphere} &&&&&&\\
			6.8 & 44.8 & 3.59 $\pm$ 0.85 & 0.62 $\pm$ 1.21 & 1.25 $\pm$ 0.40 & 5.45 $\pm$ 1.53 & \textit{High Sub-Observer}  \\
			15.3 & 27.8 & 2.76 $\pm$ 0.34 & 3.91 $\pm$ 0.38 & 0.11 $\pm$ 0.22 & 7.77 $\pm$ 0.55 & \textit{Latitudes ($\ge$ 45$\degree$):} \\
			26.7 & 48.8& 3.58 $\pm$ 0.43 & 0.42 $\pm$ 0.60 & 1.05 $\pm$ 0.23 & 5.05 $\pm$ 0.78 & 3.68 $\pm$ 1.63 \\
			\vspace{-0.0 cm}*53.6 & -16.0 & 1.36 $\pm$ 0.64 & 1.42 $\pm$ 1.19 & 0.23 $\pm$ 0.59 & 3.02 $\pm$ 1.47 & \\
			\vspace{-0.0 cm}*79.8& -19.4 & 1.34 $\pm$ 0.44 & -1.23 $\pm$ 0.56 & 0.81 $\pm$ 0.26 &  0.92 $\pm$ 0.76 & \textit{Mid Sub-Observer}  \\
			\vspace{-0.0 cm}*87.8& 24.0 & 1.26 $\pm$ 0.31 & 0.43 $\pm$ 0.44 & 0.49 $\pm$ 0.23 & 2.17 $\pm$ 0.59 & \textit{Latitudes (31 -- 44$\degree$):} \\
			\vspace{-0.0 cm}*93.5 & -18.1 & 0.87 $\pm$ 0.56 & 1.11 $\pm$ 0.72 & -0.24 $\pm$ 0.51 & 1.74 $\pm$ 1.05 & 3.12 $\pm$ 1.30  \\
			100.3 & 46.4& 1.10 $\pm$ 0.39 & 1.76 $\pm$ 0.43 & 0.42 $\pm$ 0.17 & 3.27 $\pm$ 0.60 & \\	
			110.1 & 32.0 & 1.00 $\pm$ 0.40 & 2.81 $\pm$ 0.55 & 0.41 $\pm$ 0.26 & 4.22 $\pm$ 0.73 & \textit{Low Sub-Observer} \\
			132.2 & 28.5 & 1.45 $\pm$ 0.40 & 3.10 $\pm$ 0.61 & 0.32 $\pm$ 0.28 & 4.88 $\pm$ 0.78 &\textit{Latitudes (-30 -- 30$\degree$):} \\
			144.8 & 43.2 & 0.42 $\pm$ 0.53 & 1.57 $\pm$ 0.66 & 0.04 $\pm$ 0.39 & 2.03 $\pm$ 0.93&  3.83 $\pm$ 1.03 \\
			152.9 & 51.2 & -0.12 $\pm$ 0.77 & 0.39 $\pm$ 0.97 & 0.67 $\pm$ 0.47 & 0.94 $\pm$ 1.32 & \\
			\vspace{-0.0 cm}*159.9 & -11.1 & 2.30 $\pm$ 0.53 & 2.91 $\pm$ 0.78 & 1.08 $\pm$ 0.45 & 6.29 $\pm$ 1.04 & \\
			\hline
			\textit{Trailing Hemisphere} &&&&&\\
			\vspace{-0.0 cm}*200.0 & -15.9 & 4.19 $\pm$ 0.67 & 5.94 $\pm$ 0.80 & 2.89 $\pm$ 0.45 & 13.02 $\pm$ 1.14 & \textit{High Sub-Observer} \\
			205.2 & 50.5 & 3.24 $\pm$ 0.29 & 3.10 $\pm$ 0.61 & 1.42 $\pm$ 0.16 & 7.76 $\pm$ 0.49 & \textit{Latitudes ($\ge$ 45$\degree$):} \\	
			205.5 & 46.1 & 3.72 $\pm$ 0.29 & 4.22 $\pm$ 0.46 & 1.57 $\pm$ 0.18 & 9.51 $\pm$ 0.58 & 10.06$\pm$ 0.86 \\
			\vspace{-0.0 cm}*219.8 & -17.2 & 5.77 $\pm$ 0.29 & 6.71 $\pm$ 0.68 & 2.52 $\pm$ 0.29 & 14.99 $\pm$ 0.80 & \\
			231.8 & 51.1 & 3.54 $\pm$ 0.35 & 5.10 $\pm$ 0.39 & 2.21 $\pm$ 0.23 & 10.85 $\pm$ 0.57 & \textit{Mid Sub-Observer} \\
			\vspace{-0.0 cm}*233.8 & -23.1 & 4.09 $\pm$ 0.78 & 6.31 $\pm$ 0.85 & 2.99 $\pm$ 0.45 & 13.40 $\pm$ 1.24 & \textit{Latitudes (31 -- 44$\degree$):} \\
			235.9 & 47.3 & 6.25 $\pm$ 0.67 & 3.66 $\pm$ 0.80 & 2.51 $\pm$ 0.26 & 12.42 $\pm$ 1.04 & 12.52$\pm$ 1.24 \\
			\vspace{-0.0 cm}*257.6 & -29.5 & 3.33 $\pm$ 0.50 & 6.47 $\pm$ 0.49 & 3.36 $\pm$ 0.27 & 13.16 $\pm$ 0.75 & \\
			260.2 & 47.0 & 5.54 $\pm$ 0.60 & 4.17 $\pm$ 0.79 & 1.90 $\pm$ 0.30 & 11.61 $\pm$ 1.04 & \textit{Low Sub-Observer} \\
			263.7 & 46.8 & 4.15 $\pm$ 0.27 & 4.06 $\pm$ 0.33 & 2.16 $\pm$ 0.14 & 10.37 $\pm$ 0.45 & \textit{Latitudes (-30 -- 30$\degree$):} \\
			268.3 & 39.0 & 3.63 $\pm$ 0.74 & 5.83 $\pm$ 0.80 & 3.07 $\pm$ 0.31 & 12.52 $\pm$ 1.14 & 14.15 $\pm$ 1.25 \\
			273.2 & 42.2 & 2.04 $\pm$ 1.37 & 5.49 $\pm$ 1.40 & 3.21 $\pm$ 0.66 & 10.74 $\pm$ 2.06 & \\
			\vspace{-0.0 cm}*278.3 & 24.8 & 6.06 $\pm$ 0.83 & 6.48 $\pm$ 1.02 & 2.83 $\pm$ 0.60 & 15.37 $\pm$ 1.45 & \\
			292.8 & 51.2 & 2.68 $\pm$ 0.41 & 3.64 $\pm$ 0.57 & 1.85 $\pm$ 0.29 & 8.16 $\pm$ 0.76 & \\		
			294.8 & 39.5 & 4.49  $\pm$ 0.36 & 6.91 $\pm$ 0.56 & 2.90 $\pm$ 0.25 & 14.30 $\pm$ 0.71 & \\
			\vspace{-0.0 cm}*294.8 & -19.3 & 6.26 $\pm$ 0.85 & 5.56 $\pm$ 0.47 & 3.08 $\pm$ 0.25 & 14.90 $\pm$ 0.66 & \\
			\vspace{-0.0 cm}*304.8 & -23.2 & 5.12 $\pm$ 0.76 & 6.29 $\pm$ 0.96 & 3.35 $\pm$ 0.61 & 14.76 $\pm$ 1.36 & \\
			\vspace{-0.0 cm}*316.6 & -18.2 & 4.52 $\pm$ 0.58 & 6.72 $\pm$ 0.82 & 2.32 $\pm$ 0.42 & 13.56 $\pm$ 1.11 & \\
			\hline
		\end{tabular}
		$^\dagger$\textit{Values shown graphically in Figure 2a.} \\
		$^\dagger$$^\dagger$\textit{Values shown graphically in Figure 2b.} \\
		\vspace{-0.0 cm}*\textit{Reported previously in \citep[]{grundy2003discovery,grundy2006distributions,cartwright2015distribution}.}
	\end{table}
	
		\begin{figure}[t!]
		\hspace{0.7 cm}\includegraphics[scale=0.85]{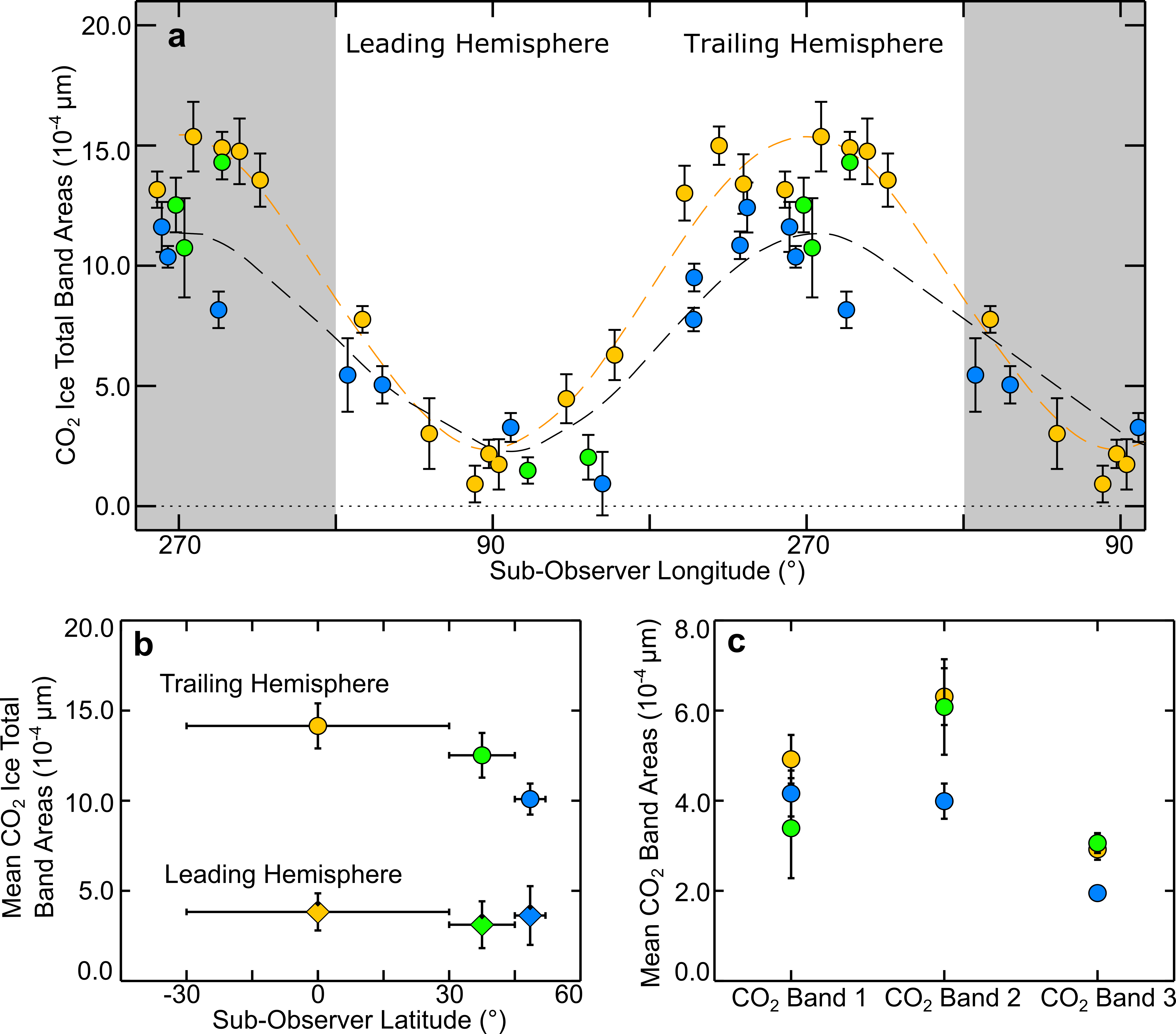}
		\caption{\textit{(a) CO$_2$ ice total band areas and 1$\sigma$ uncertainties for Ariel's low  (30$\degree$S -- 30$\degree$N), mid (31$\degree$ -- 44$\degree$N), and high (45$\degree$ -- 51$\degree$N) sub-observer latitude zones (gold, green, and blue dots, respectively). Numerical values for these data points are shown in the `CO$_2$ Total Area' column in Table 2. A sinusoidal fit to the low sub-observer latitude data points (gold dashed line) highlights the clear longitudinal trend in the distribution of CO$_2$ ice on Ariel in this latitude zone. Similarly, the sinusoidal fit to the mid and high sub-observer latitude data points (black dashed line) highlights the reduction in CO$_2$ ice band areas on Ariel's trailing hemisphere at sub-observer latitudes $\ge$ 30$\degree$N. Repeat longitudes (gray-toned zones) are shown to highlight the periodicity in the distribution of CO$_2$ on Ariel. (b) Mean CO$_2$ ice total band areas and 1$\sigma$ uncertainties (vertical error bars) for the low, mid, and high sub-observer latitude zones (gold, green, and blue symbols, respectively) over Ariel's leading (diamonds) and trailing (circles) hemispheres. The horizontal error bars represent the range of latitudes we use to define each sub-observer latitude zone.  Numerical values for these data points are shown in the `Mean CO$_2$ Total Areas' column in Table 2. (c) Mean CO$_2$ band 1, 2, and 3 areas and 1$\sigma$ uncertainties for the low, mid, and high sub-observer latitude zones (gold, green, and blue dots, respectively) on Ariel's trailing hemisphere.}}\vspace{0.1 cm}
	\end{figure}

	\subsection{Distribution of CO$_2$ ice on Ariel} 
	The 13 Ariel spectra analyzed previously demonstrate that CO$_2$ bands are much stronger in the spectra collected over Ariel's trailing hemisphere compared to its leading hemisphere \citep[]{grundy2006distributions,cartwright2015distribution}. Utilizing the same measurement procedure presented in \citet{cartwright2015distribution} for the analysis of those 13 previously analyzed Ariel spectra, we assessed the distribution of CO$_2$ ice in the 18 previously unanalyzed Ariel spectra (Figure 2a). The measurements show similar trends in the longitudinal distribution of CO$_2$ ice in the spectra collected over low, mid, and high sub-observer latitudes. However, the previously determined leading/trailing hemispherical dichotomy in the distribution of CO$_2$ ice appears to be weaker for the spectra collected at mid and high sub-observer latitudes compared to low sub-observer latitudes (Table 2, Figure 2a). 
	
	We then calculated the percentage of Ariel's observed disk that was composed of low latitude regions (30$\degree$S $-$ 30$\degree$N) for all of the spectra collected over its trailing hemisphere (see \citealt{holler2016surface} for disk-area calculation details). We found that for spectra where the disk of Ariel was composed of $>$ 55$\%$ low latitude regions, the total band areas for detected CO$_2$ ice features ranges from 13.02 $\pm$ 1.14 $\cdot$ 10$^-$$^4$$\micron$ to 15.37 $\pm$ 1.45 $\cdot$ 10$^-$$^4$$\micron$. When Ariel's disk was composed of 45 -- 55$\%$ low latitude regions, the measured CO$_2$ band areas decreased to 10.74 $\pm$ 2.06 $\cdot$ 10$^-$$^4$$\micron$ to 14.30 $\pm$ 0.71 $\cdot$ 10$^-$$^4$$\micron$. When low latitudes regions composed $<$ 45$\%$ of Ariel's observed disk, CO$_2$ band areas decreased further to 7.76 $\pm$ 0.49 $\cdot$ 10$^-$$^4$$\micron$ to 12.42 $\pm$ 1.04 $\cdot$ 10$^-$$^4$$\micron$. Thus, CO$_2$ ice band areas are gradually decreasing as the sub-observer latitude moves further north and away from Ariel's low latitudes.
	
	To further investigate latitudinal trends in the distribution of CO$_2$ ice, we compared the mean CO$_2$ ice total band areas of each sub-observer latitude zone (final column of Table 2, Figure 2b). This comparison shows that the low sub-observer latitude zone has the largest mean CO$_2$ total band area. The mid sub-observer latitude zone has a smaller mean CO$_2$ total band area, and the high sub-observer latitude zone has the smallest mean CO$_2$ total band area ($>$ 1$\sigma$ difference between CO$_2$ band areas for the low and high sub-observer latitude zones). On Ariel's leading hemisphere, these mean measurements show negligible differences between the three sub-observer latitude zones ($<$ 1$\sigma$ difference). 
	
	We also calculated the mean areas for each of the three CO$_2$ bands (Figure 2c). CO$_2$ band 1 shows $<$ 1$\sigma$ difference between the different sub-observer latitude zones, whereas CO$_2$ bands 2 and 3 show $>$ 2$\sigma$ differences between the high and low sub-observer latitude zones ($>$ 1$\sigma$ difference between the high and mid sub-observer latitude zones). Of note, CO$_2$ band 1 is located on the long wavelength end of a strong telluric band resulting from H$_2$O vapor, possibly explaining the inconclusive results for this band compared to CO$_2$ bands 2 and 3, which are less contaminated by species in Earth's atmosphere. Thus, our measurements demonstrate that CO$_2$ ice bands are stronger in spectra collected at low and mid sub-observer latitudes compared to spectra collected at high sub-observer latitudes over Ariel's trailing hemisphere. Spectra collected over Ariel's leading hemisphere do not display discernible differences in CO$_2$ band areas at different sub-observer latitudes. 
	
	\subsection{Radiative transfer modeling of CO$_2$ ice bands} 

	\begin{figure}[h]
	\includegraphics[scale=1.04]{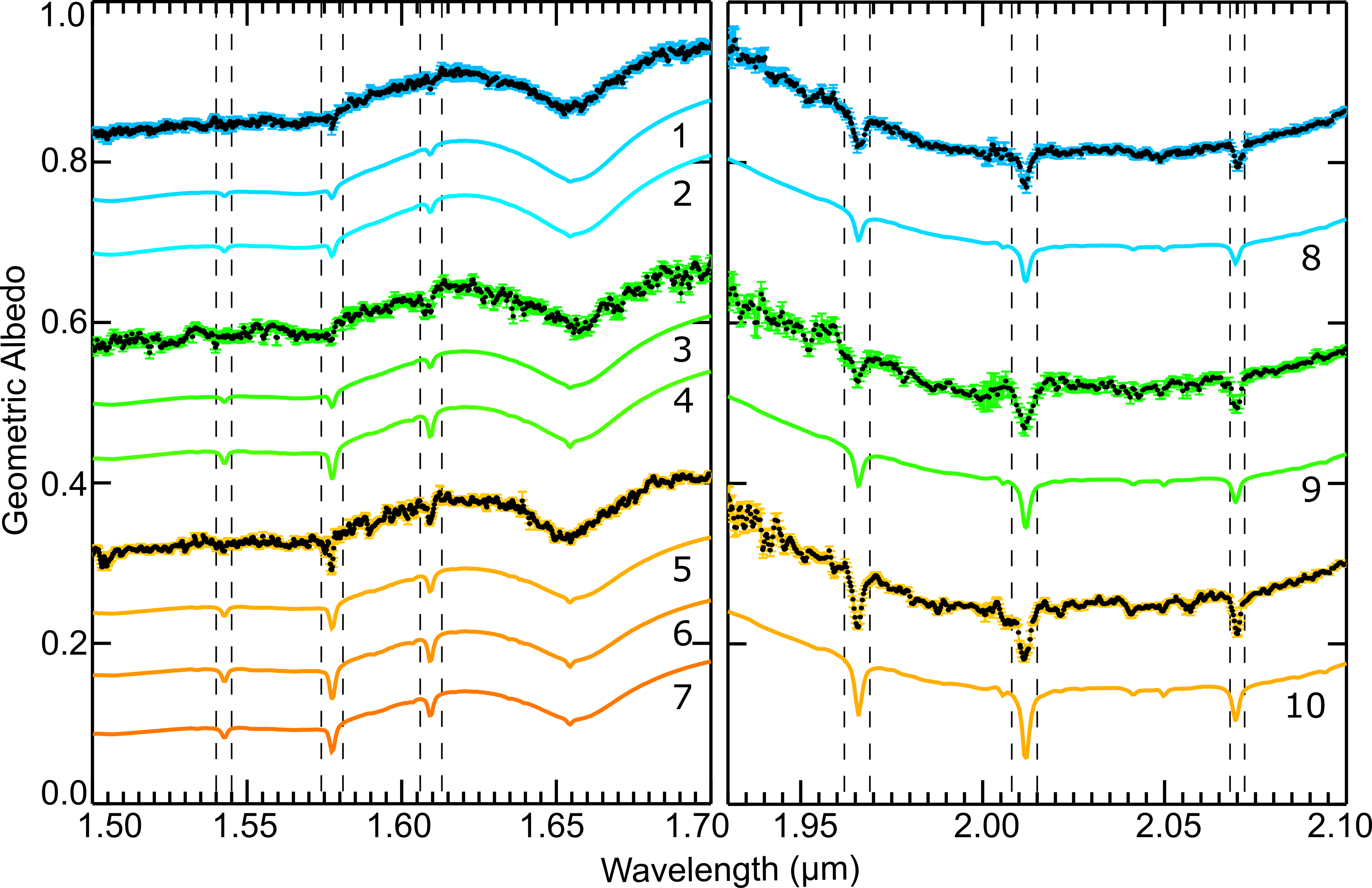}
	\caption{\textit{Left: Comparison between the grand average Ariel spectra for low, mid, and high latitude zones (1$\sigma$ uncertainties indicated by gold, green, and blue error bars, respectively) and synthetic spectra (numbered 1 -- 7) that model the CO$_2$ ice combination and overtone bands centered near 1.543, 1.578, and 1.609 $\micron$ (CO$_2$ bands bracketed by black dashed lines). Right: Comparison between the grand average Ariel spectra for low, mid, and high latitude zones (orange, green, and blue error bars, respectively) and synthetic spectra (numbered 8 -- 10) that model the CO$_2$ ice combination and overtone bands centered near 1.966, 2.012, and 2.070 $\micron$ (CO$_2$ bands bracketed by black dashed lines). The low latitude synthetic spectrum (10) was originally reported in \citet[]{cartwright2015distribution}. The included constituents for each synthetic spectrum are summarized in Tables 3 and 4. Spectra and models in both plots are scaled to Ariel's geometric albedo (0.564) at $\sim$0.958 $\micron$ \citep[]{karkoschka2001comprehensive} and offset vertically for clarity. The grand average spectra were lightly smoothed using a 2 to 4 pixel wide boxcar function.}} \vspace{0.1 cm}
\end{figure}
	
		\begin{table}[hbt!]
		\caption {Model parameters for synthetic spectra: CO$_2$ ice bands between 1.5 and 1.7 $\micron$.} 
		\vspace{-0.1 cm}\hskip-1.4 cm\begin{tabular}{cc|cccc|cc}
			\hline\hline
			& \multicolumn{5}{c}{Particulate Mixture Components} & \multicolumn{2}{c}{Areal Mixture Components} \\
			\begin{tabular}[c]{@{}l@{}}\hspace{-1 cm}  $^\dagger$Synthetic \\ \hspace{-1 cm} Spectrum \end{tabular} & \begin{tabular}[c]{@{}l@{}} \hspace{-1 cm}CO$_2$ Ice \\ \hspace{-1 cm}Band \end{tabular} & \begin{tabular}[c]{@{}l@{}} \hspace{-1 cm} H$_2$O ice: \\ \hspace{-1 cm} Grain Size, \\ \hspace{-1 cm} Fractional \\ \hspace{-1 cm} Coverage \end{tabular} & \begin{tabular}[c]{@{}l@{}}  \hspace{-1 cm} H$_2$O ice: \\ \hspace{-1 cm} Grain Size, \\ \hspace{-1 cm} Fractional \\ \hspace{-1 cm} Coverage \end{tabular} & \begin{tabular}[c]{@{}l@{}}  \hspace{-1 cm} H$_2$O ice: \\ \hspace{-1 cm} Grain Size, \\ \hspace{-1 cm} Fractional \\ \hspace{-1 cm} Coverage \end{tabular} & \begin{tabular}[c]{@{}l@{}}  \hspace{-1 cm} Amorphous C: \\ \hspace{-1 cm} Grain Size, \\ \hspace{-1 cm} Fractional \\ \hspace{-1 cm} Coverage \end{tabular} & \begin{tabular}[c]{@{}l@{}}  \hspace{-1 cm} CO$_2$ ice: \\ \hspace{-1 cm} Grain Size, \\ \hspace{-1 cm} Fractional \\ \hspace{-1 cm} Coverage \end{tabular} & \begin{tabular}[c]{@{}l@{}}  \hspace{-1 cm} CO$_2$ ice: \\ \hspace{-1 cm} Grain Size, \\ \hspace{-1 cm} Fractional \\ \hspace{-1 cm} Coverage \end{tabular} \\
			\hline
			\textit{High Sub-Obs. Lat.} & & & & & & & \\		
			1 & 1.58 $\micron$ &50 $\micron$, & 10 $\micron$, & 0.3 $\micron$, & 12.5 $\micron$, & 50 $\micron$, & 10 $\micron$,  \\
			&  &26.8$\%$ & 57.0$\%$ & 0.3$\%$ & 0.9$\%$  & 7.5$\%$ & 7.5$\%$  \\	
			2 & 1.61 $\micron$ &50 $\micron$, & 10 $\micron$, & 0.3 $\micron$, & 12.5 $\micron$, & 100 $\micron$, & 50 $\micron$,  \\
			&  &26.5$\%$ & 56.3$\%$ & 0.3$\%$ & 0.9$\%$  & 8.8$\%$ & 7.2$\%$  \\			
			\textit{Mid Sub-Obs. Lat.} & & & & & & & \\		
			3  & 1.58 $\micron$ &50 $\micron$, & 10 $\micron$, & 0.3 $\micron$, & 12.5 $\micron$, & 50 $\micron$, & 10 $\micron$,  \\
			& &55.1$\%$ & 20.8$\%$ & 0.3$\%$ & 0.8$\%$  & 13.8$\%$ & 9.2$\%$  \\
			4 & 1.61 $\micron$ &50 $\micron$, & 10 $\micron$, & 0.3 $\micron$, & 12.5 $\micron$, & 200 $\micron$, & 50 $\micron$,  \\
			& &54.3$\%$ & 20.6$\%$ & 0.3$\%$ & 0.8$\%$  & 21.6$\%$ & 2.4$\%$  \\		
			\textit{Low Sub-Obs. Lat.} & & & & & & & \\		
			5 & 1.54 $\micron$ &50 $\micron$, & 10 $\micron$, & 0.3 $\micron$, & 12.5 $\micron$, & 50 $\micron$, & 10 $\micron$, \\
			& &51.6$\%$ & 11.9$\%$ & 0.3$\%$ & 0.7$\%$  & 8.9$\%$ & 26.6$\%$ \\
			6 & 1.58 $\micron$ &50 $\micron$, & 10 $\micron$, & 0.3 $\micron$, & 12.5 $\micron$, & 200 $\micron$, & 50 $\micron$, \\
			& &50.8$\%$ & 11.7$\%$ & 0.3$\%$ & 0.7$\%$  & 7.3$\%$ & 29.2$\%$ \\
			7 & 1.61 $\micron$ &50 $\micron$, & 10 $\micron$, & 0.3 $\micron$, & 12.5 $\micron$, & 100 $\micron$, & 10 $\micron$, \\
			& &49.2$\%$ & 11.4$\%$ & 0.2$\%$ & 0.7$\%$  & 21.2$\%$ & 17.3$\%$ \\
			\hline
		\end{tabular}
		\vspace{-0.6 cm}\hspace{0.5 cm}$^\dagger$\textit{Synthetic spectra shown graphically in Figure 3.} \\
	\end{table}

	In Figure 3, we present ten spectral models that fit the CO$_2$ ice bands detected on Ariel's trailing hemisphere between 1.5 and 1.7 $\micron$ (synthetic spectra 1 -- 7) and 1.9 and 2.1 $\micron$ (synthetic spectra 8 -- 10). The fractional coverage and grain sizes of all included components are summarized in Table 3 (CO$_2$ ice bands centered near 1.543, 1.578, and 1.609 $\micron$) and Table 4 (CO$_2$ bands 1, 2, and 3). The previously reported spectral models for CO$_2$ bands 1, 2, and 3 in Ariel's low sub-observer latitude, grand average spectrum \citep[]{cartwright2015distribution} are included here (synthetic spectrum 10) along with new models of the mid and high sub-observer latitude, grand average spectra (synthetic spectra 9 and 10, respectively). The low sub-observer latitude synthetic spectrum includes 27$\%$ CO$_2$ ice, the mid sub-observer latitude model includes 23$\%$ CO$_2$ ice, and the high sub-observer latitude model includes 16$\%$ CO$_2$ ice (model components are summarized in Table 4). 

	 The low sub-observer latitude, grand average spectrum shows the strongest evidence for the 1.543-$\micron$, 1.578-$\micron$, and 1.609-$\micron$ CO$_2$ bands, which we fit with three different models (synthetic spectra 5 -- 7) spanning 1.5 to 1.7 $\micron$. The amount of CO$_2$ ice included in synthetic spectra 5 -- 7 ranges between 35.5$\%$ to 38.5$\%$, which is substantially higher than the amount of CO$_2$ included in synthetic spectrum 10 (27$\%$), spanning 1.9 to 2.1 $\micron$. Furthermore, synthetic spectra 5 -- 7 include greater amounts of larger CO$_2$ ice grains, including grains with diameters of 100 and 200 $\micron$, which are not included in any of the models spanning 1.9 to 2.1 $\micron$ (synthetic spectra 8 -- 10).

		\begin{table}[hbt!]
		\caption {Model parameters for synthetic spectra: CO$_2$ bands 1, 2, and 3.} 
		\vspace{-0.1 cm}\hskip-1.2 cm\begin{tabular}{*5c|cc}
			\hline\hline
			& \multicolumn{4}{c}{Particulate Mixture Components} & \multicolumn{2}{c}{Areal Mixture Components} \\
			\begin{tabular}[c]{@{}l@{}}\hspace{-1 cm} $^\dagger$Synthetic \\ \hspace{-1 cm} Spectrum \end{tabular} & \begin{tabular}[c]{@{}l@{}} \hspace{-1 cm} H$_2$O ice: \\ \hspace{-1 cm} Grain Size, \\ \hspace{-1 cm} Fractional \\ \hspace{-1 cm} Coverage \end{tabular} & \begin{tabular}[c]{@{}l@{}}  \hspace{-1 cm} H$_2$O ice: \\ \hspace{-1 cm} Grain Size, \\ \hspace{-1 cm} Fractional \\ \hspace{-1 cm} Coverage \end{tabular} & \begin{tabular}[c]{@{}l@{}}  \hspace{-1 cm} H$_2$O ice: \\ \hspace{-1 cm} Grain Size, \\ \hspace{-1 cm} Fractional \\ \hspace{-1 cm} Coverage \end{tabular} & \begin{tabular}[c]{@{}l@{}}  \hspace{-1 cm} Amorphous C: \\ \hspace{-1 cm} Grain Size, \\ \hspace{-1 cm} Fractional \\ \hspace{-1 cm} Coverage \end{tabular} & \begin{tabular}[c]{@{}l@{}}  \hspace{-1 cm} CO$_2$ ice: \\ \hspace{-1 cm} Grain Size, \\ \hspace{-1 cm} Fractional \\ \hspace{-1 cm} Coverage \end{tabular} & \begin{tabular}[c]{@{}l@{}}  \hspace{-1 cm} CO$_2$ ice: \\ \hspace{-1 cm} Grain Size, \\ \hspace{-1 cm} Fractional \\ \hspace{-1 cm} Coverage \end{tabular} \\
			\hline
			\textit{High Sub-Obs. Lat.} & & & & & & \\
			8 &  50 $\micron$, & 10 $\micron$, & 0.3 $\micron$, & 12.5 $\micron$, & 50 $\micron$, & 1 $\micron$,  \\
			& 25.2$\%$ & 56.3$\%$ & 0.7$\%$ & 1.8$\%$  & 9.6$\%$ & 6.4$\%$  \\
			\textit{Mid Sub-Obs. Lat.} & & & & & & \\
			9 &  50 $\micron$, & 10 $\micron$, & 0.3 $\micron$, & 12.5 $\micron$, & 50 $\micron$, & 1 $\micron$,  \\
			& 54.6$\%$ & 21.1$\%$ & 0.6$\%$ & 1.7$\%$  & 7.7$\%$ & 14.3$\%$  \\
			\textit{Low Sub-Obs. Lat.} & & & & & & \\
			\vspace{-0. cm}*10 &  50 $\micron$, & 10 $\micron$, & 0.3 $\micron$, & 12.5 $\micron$, & 50 $\micron$, & 10 $\micron$, \\
			& 36.5$\%$ & 34.3$\%$ & 0.6$\%$ & 1.6$\%$  & 22.5$\%$ & 4.5$\%$ \\		
			\hline
	\end{tabular}
		\\
		\hspace{3 cm}$^\dagger$\textit{Synthetic spectra shown graphically in Figure 3.} \\
     	\hspace{1.2 cm}*\textit{Previously reported in \citep[]{cartwright2015distribution}.}
	\end{table}

	\begin{table}[hbt!]
      \caption {$\chi_R^2$ Statistics for synthetic spectra.} 
    \vspace{-0.1cm} \hskip-1.2cm \begin{tabular}{*5c|*4c}
		\hline\hline
		\begin{tabular}[c]{@{}l@{}}\hspace{-1cm}  \\ \hspace{-1cm} \end{tabular} & 
		\begin{tabular}[c]{@{}l@{}} \hspace{-1cm}1.543-$\micron$  \\ \hspace{-1cm}CO$_2$ Ice \\ \hspace{-1cm}Band \end{tabular} & 
		\begin{tabular}[c]{@{}l@{}} \hspace{-1cm}1.578-$\micron$  \\ \hspace{-1cm}CO$_2$ Ice \\ \hspace{-1cm}Band \end{tabular} & 		
		\begin{tabular}[c]{@{}l@{}} \hspace{-1cm}1.609-$\micron$  \\
	    \hspace{-1cm}CO$_2$ Ice \\ \hspace{-1cm}Band \end{tabular} & 
        \begin{tabular}[c]{@{}l@{}} \hspace{-1cm}Continuum  \\ \\ \end{tabular} & 
		\begin{tabular}[c]{@{}l@{}} \hspace{-1cm}1.966-$\micron$  \\ \hspace{-1cm}CO$_2$ Ice \\ \hspace{-1cm}Band \end{tabular} & 
		\begin{tabular}[c]{@{}l@{}} \hspace{-1cm}2.012-$\micron$  \\ \hspace{-1cm}CO$_2$ Ice \\ \hspace{-1cm}Band \end{tabular} & 
		\begin{tabular}[c]{@{}l@{}} \hspace{-1cm}2.070-$\micron$  \\ \hspace{-1cm}CO$_2$ Ice \\ \hspace{-1cm}Band \end{tabular}  &	
		\begin{tabular}[c]{@{}l@{}} \hspace{-1cm}Continuum  \\ \\ \end{tabular} \\						
		\hline
	    Wavelength & 1.540 -- & 1.574 -- & 1.606 -- & 1.5 -- 1.7 & 1.962 -- & 2.008 -- & 2.068 -- & 1.9 -- 2.1  \\
	    Range ($\micron$) & 1.545 & 1.581 & 1.613 & & 1.969 & 2.015 & 2.072 &  \\
		\hline	
		\textit{Synthetic} & &  & & & & & & \\  
		\textit{Spectra} & &  & & & & & & \\      	
		1 & - & \textbf{0.789} & 0.993 & 0.029 & - & - & - & -  \\
	    2 & - & 0.849  &\textbf{1.004} & 0.029 & - & - & - & -  \\
		3 & - & \textbf{0.894} &0.876  & 0.163 & - & - & - \\
		4 & - & 1.600  & \textbf{0.847} & 0.182 & -  & - & - \\
		5 &  \textbf{0.958} & 0.709 & 1.365 & 0.065 & - & - & - \\
		6 & 1.117 & \textbf{0.836} & 1.512 & 0.064 & - & - & - \\
		7 & 6.927 & 0.717 & \textbf{0.386} & 0.111 & - & - & - \\	
		8 & - & - & - & - & 0.227 & 0.225 & 1.028 & 0.126 \\	
		9 & - & - & - & - & 0.434 & 0.677 & 0.567 & 0.348 \\	
		10 & - & - & - & - & 0.366 & 1.286 & 0.667 & 0.688 \\							
		\hline
	\end{tabular}
		\\ \\
		\hspace{0cm}\textit{Bolded values highlight the primary CO$_2$ ice band fit by each of the synthetic spectra spanning 1.5 to 1.7 $\micron$.} \\
    \end{table}

	In contrast, Ariel's grand average mid and high sub-observer latitude, grand average spectra display weaker 1.578-$\micron$ and 1.609-$\micron$ CO$_2$ bands compared to the low sub-observer latitude spectrum, and the weak 1.543-$\micron$ band is not visibly identifiable in either spectrum (Figure 3). The spectral models for Ariel's mid sub-observer latitude spectrum that we present here (synthetic spectra 3 and 4) include 23 and 24$\%$ CO$_2$ to fit the 1.578-$\micron$ and 1.609-$\micron$ bands, respectively. These two synthetic spectra include greater amounts of larger grains (10, 50, and 200 $\micron$ diameters) compared to synthetic spectrum 9 that spans 1.9 -- 2.1 $\micron$ of the mid sub-observer latitude, grand average spectrum (1 and 50 $\micron$ diameters). Similarly, the spectral models for Ariel's high sub-observer latitude, grand average spectrum (synthetic spectra 1 and 2) include 15 and 16$\%$ CO$_2$ to fit the 1.578-$\micron$ and 1.609-$\micron$ bands, respectively. Synthetic spectra 1 and 2 also include greater amounts of larger grains (10, 50, and 100 $\micron$ diameters) compared to synthetic spectrum 8 spanning 1.9 to 2.1 $\micron$ (1 and 50 $\micron$ diameters). Therefore, all of the synthetic spectra spanning 1.5 to 1.7 $\micron$ include more CO$_2$ ice, and larger CO$_2$ ice grains, than the corresponding synthetic spectra that span 1.9 to 2.1 $\micron$.
	
	Along with visual assessment, we utilized $\chi_R^2$ tests to assess the quality of the fits provided by the synthetic spectra. The $\chi_R^2$ statistics for the 1.543-$\micron$ CO$_2$ ice band (synthetic spectra 5 -- 7) and the 1.578-$\micron$ and 1.609-$\micron$ CO$_2$ bands (synthetic spectra 1 -- 7) demonstrate that the generated models are suitable fits to the observed spectra (Table 5). Similarly, the $\chi_R^2$ values for CO$_2$ ice bands 1, 2, and 3 (synthetic spectra 8 -- 10) demonstrate that the generated models are suitable fits to the observed spectra. The $\chi_R^2$ values for the synthetic spectra spanning the 1.5 to 1.7 $\micron$ and 1.9 to 2.1 $\micron$ continua also appear to provide reasonable fits, albeit some of the $\chi_R^2$ values for the synthetic spectra spanning 1.5 to 1.7 $\micron$ are quite low ($<$ 0.1) and may result from overestimated errors. Comparison of the $\chi_R^2$ statistics for synthetic spectra 1 -- 7 suggests that the small differences in the grain size ranges and the amount of CO$_2$ ice included in each of these spectral models are not sufficient to cause significant variations in their goodness-of-fit for each of the CO$_2$ ice bands between 1.5 and 1.7 $\micron$ (Table 5). More sophisticated and computationally intensive model assessment techniques may be necessary to better delineate between the fits provided by synthetic spectra 1 -- 7. Nevertheless, the small differences between each of these synthetic spectra may reflect subtle spectral variations resulting from differences in photon penetration depths as a function of wavelength (see Section 4.2).  
		
	In summary, the synthetic spectra we generated for CO$_2$ ice bands between 1.5 and 1.7 $\micron$ and 1.9 and 2.1 $\micron$ suggest that the fractional coverage of CO$_2$ ice is highest at low sub-observer latitudes, decreases at mid sub-observer latitudes, and decreases further at high sub-observer latitudes. These spectral modeling  results are consistent with our CO$_2$ band area measurements, indicating that CO$_2$ ice is likely concentrated at low latitudes on Ariel. 
	
	
	\section{Discussion} 
	
	\subsection{Nature and distribution of CO$_2$ ice on Ariel}  
	Thermodynamical models predict that CO$_2$ molecules generated and/or exposed at polar latitudes on the surfaces of the Uranian moons should readily sublimate during summer when the pole is continuously exposed to direct sunlight \citep{grundy2006distributions,sori2017wunda}.  In the high-obliquity Uranian system, equatorial regions experience less net sublimation than polar regions over the course of a Uranian year. Consequently, CO$_2$ should migrate over geologically short timescales from these moons' poles to their equators, thereby enriching their low latitudes with CO$_2$ ice. If CO$_2$ molecules are distributed uniformly across the surfaces of the Uranian moons, then CO$_2$ would migrate to cold traps at latitudes between 50$\degree$S and 50$\degree$N, with the removal of CO$_{2}$ increasing steadily with latitude, peaking near their poles (see Fig. 9 in \citealt{sori2017wunda}). Conversely, if CO$_2$ is already depleted from latitudes $>$ 50$\degree$ S or N, then the zone of CO$_2$ accumulation may be smaller, spanning only 30$\degree$S to 30$\degree$N, with peak CO$_2$ removal concentrated near 45$\degree$S and 45$\degree$N. In both model scenarios, CO$_2$ molecules get cold trapped at similar longitudes to where they were initially generated and/or exposed. Thus, the apparent concentration of CO$_2$ ice on the trailing hemispheres of Ariel and the other Uranian moons suggests that CO$_2$ molecules originate on their trailing sides \citep{sori2017wunda}. 
	
	Our band area measurements and synthetic spectra are consistent with a reduction in the abundance of CO$_2$ ice at higher latitudes on Ariel. However, the disk-integrated nature of the data analyzed here limits our ability to develop more detailed maps of the latitudinal distribution of CO$_2$ ice. Our analyses are therefore unable to discern whether the abundance of CO$_2$ ice is steadily decreasing with increasing sub-observer latitude, or whether CO$_2$ ice is mostly concentrated in one or perhaps a few large cold traps at low latitudes and essentially absent at latitudes $\ge$ 30$\degree$N. 
	
	Other planetary bodies display large and bright volatile-rich regions, such as the H$_2$O and CO$_2$ ice dominated Martian poles \citep[e.g.,][]{farmer1976mars,kieffer1979mars,kieffer2000mars} and the N$_2$ ice and CO ice dominated composition of Sputnik Planita at low northern latitudes on Pluto \citep[e.g.,][]{bertrand2016observed,hamilton2016rapid,earle2017long}. Furthermore, the available photogeologic evidence \citep[e.g.,][]{smith1986voyager} and photometric measurements made with the Voyager 2 Imaging Science System \citep[e.g.,][]{buratti1991comparative} indicate that a large and bright annulus of material is mantling the floor of Wunda crater on Umbriel's trailing hemisphere, which has been interpreted to be a cold trap for CO$_2$ ice \citep[]{sori2017wunda}. Ariel, on the other hand, does not appear to display large and bright regions on its trailing hemisphere that might serve as volatile sinks for CO$_2$ ice. Nevertheless, large CO$_2$ ice cold traps could be present at low northern latitudes, which were shrouded by winter darkness during the Voyager 2 flyby and were not imaged. Continued observations of Ariel as the Uranian system migrates into northern summer in February 2030, reaching a maximum sub-solar latitude of $\sim$82$\degree$N, could be used to help discern between the two different modeling predictions for the latitudinal distribution of CO$_2$ ice presented in \citet{sori2017wunda}.
	
	\textit{Radiolytic production of CO$_2$ molecules?} Charged particle bombardment of H$_2$O ice grains mixed with C-rich material in Ariel's regolith could drive radiolytic production of oxidized carbon-based species such as CO$_2$. Radiolytically generated CO$_2$ molecules should not be thermodynamically stable in Ariel's dark regolith, in particular at high latitudes where the timescales of maximum solar heating are considerably longer than at low latitudes \citep[]{grundy2006distributions,sori2017wunda}. Consequently, CO$_2$ molecules radiolytically generated in Ariel's polar regions should diffuse out of its regolith, hop along its surface, and get cold trapped in its equatorial region over seasonal timescales. Assuming the source(s) of CO$_2$ molecules operate over long timescales, then low latitude cold traps should gradually expand as they accumulate more CO$_2$, forming an optically thick layer of CO$_2$ ice, with a crystalline structure dominated by CO$_2$ molecules (i.e., a pure CO$_2$ ice deposit). The detection of the weak CO$_2$ ice band between 1.5 and 1.7 $\micron$, which are a factor of 60 to 200 weaker than CO$_2$ ice bands 1, 2, and 3 \citep[]{hansen1997spectral}, supports the interpretation that thick deposits of CO$_2$ ice are present on Ariel. Once trapped in CO$_2$ ice deposits, CO$_2$ molecules should be fairly stable and resistant to sublimation due to the deposit's higher albedo and thermal inertia compared to the surrounding regolith \citep{sori2017wunda}. 
	
	Along with producing new CO$_2$ molecules, interactions with energetic electrons (\textit{e$^-$}) and protons (\textit{p$^+$}) should decompose CO$_2$ molecules, forming CO \citep[e.g.,][]{gerakines2001energetic,zheng2007formation}:
	
	\begin{center}\textit{e$^-$} + CO$_2$ $\rightarrow$ CO + O\\
							 \textit{p$^+$} + CO$_2$ $\rightarrow$ CO + O,\end{center}	

	CO and O are likely transient and should back-react with surrounding species to form new radiolytic products. As a result, CO and O fragments generated by irradiation of pure CO$_2$ ice deposits should primarily recombine into CO$_2$, possibly enhancing the persistence of CO$_2$ ice deposits on the surface of Ariel and the other Uranian moons. Supporting this scenario, some spectra collected over Ariel's trailing hemisphere exhibit a subtle band centered near 2.35 $\micron$ that could result from CO \citep[e.g,][]{gerakines2001energetic,zheng2007formation}, in particular spectra collected over low sub-observer latitudes where CO$_2$ ice bands are strongest (Figure 4). This 2.35-$\micron$ band does not appear to be present in spectra collected over Ariel's leading hemisphere, where CO$_2$ ice is essentially absent. CO and O fragments are volatile at Ariel's estimated peak surface temperature (84 $\pm$ 1 K, \citealt{hanel1986infrared}) and should be rapidly removed if exposed on Ariel's surface. Fragments of CO$_2$ molecules dissociated at depth within CO$_2$ ice deposits, however, might be retained long enough for some of the newly formed CO to back-react and re-form CO$_2$. Additionally, perhaps some CO and O fragments get trapped in void spaces and defects within the crystalline structure of Ariel's CO$_2$ deposits instead of back-reacting or escaping. In this scenario, O fragments would likely combine into more stable O$_2$ molecules before getting trapped. A possibly analogous process is suspected to occur on the icy Galilean moons, which display a subtle 0.577-$\micron$ absorption band attributed to trapped O$_2$ formed by radiolysis of H$_2$O ice on their surfaces, which is subsequently trapped in defects and voids in the crystalline structure of H$_2$O ice \citep[e.g.,][]{spencer1995charge,spencer2002condensed}. 

\begin{figure}[h!]
	\includegraphics[scale=0.75]{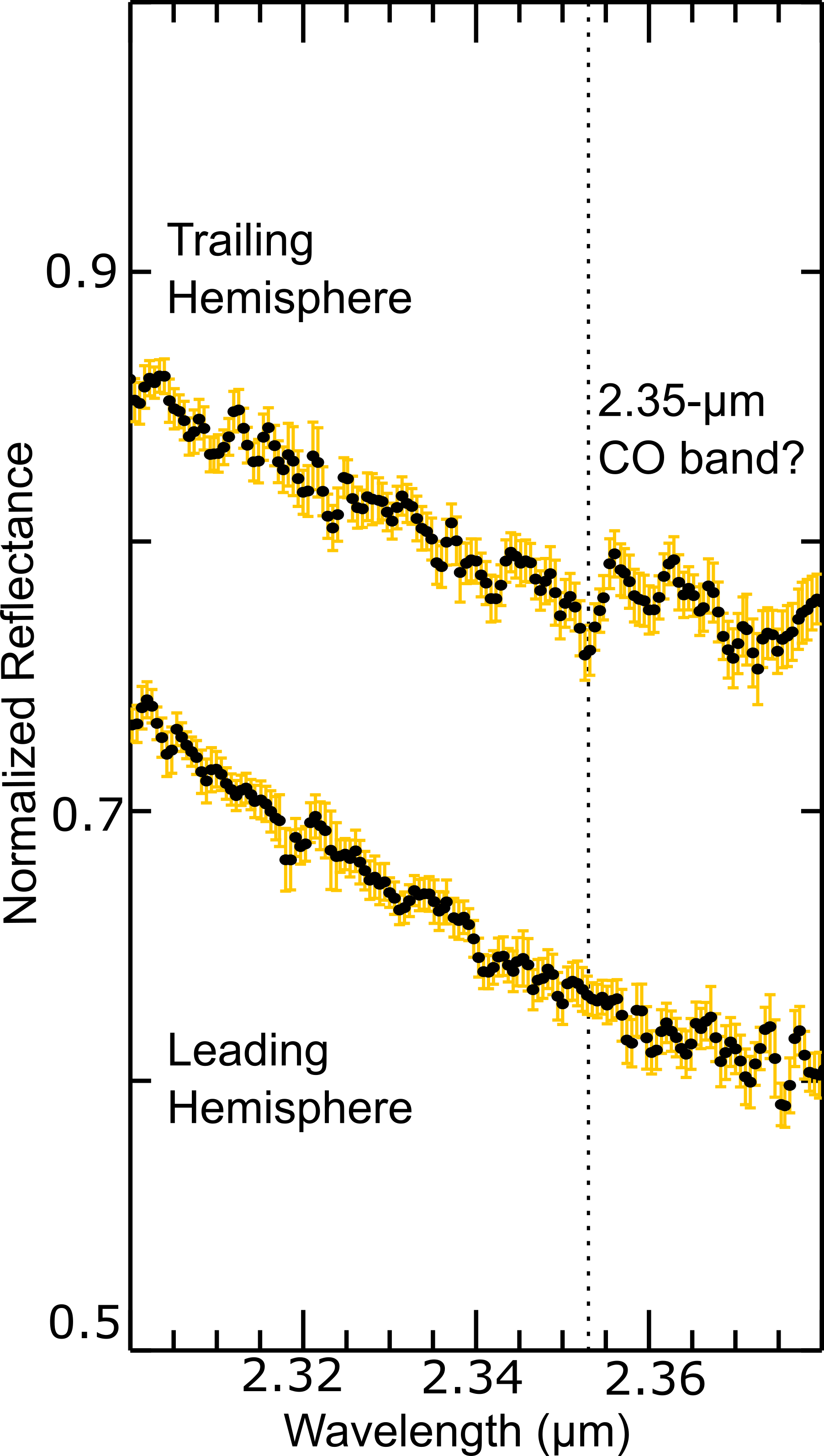}
	\caption{\textit{Grand average low latitude spectra and 1$\sigma$ uncertainties (gold) of Ariel's trailing (top) and leading (bottom) hemisphere, highlighting the position of a subtle 2.35 $\micron$ band (black dotted line) that is observed on Ariel's trailing hemisphere but is absent from its leading hemisphere. Both spectra were normalized to 1 using the mean reflectance between 1.74 and 1.77 $\micron$ and offset vertically for clarity. These data were lightly smoothed using a 5-pixel wide boxcar function. }}\vspace{1 cm}
\end{figure} 
	
\textit{Regolith mixed CO$_2$ ice on Ariel?} The spectral signature of CO$_2$ on the Uranian moons is consistent with crystalline CO$_2$ ice, segregated from other constituents in concentrated deposits. However, if CO$_2$ molecules are being actively generated by charged particle bombardment of H$_2$O ice grains and C-rich material in Ariel's regolith, then the spectral signature of these radiolytic production sites might be consistent with CO$_2$ well mixed with H$_2$O and other surface constituents. Laboratory experiments that investigated the spectral signature of CO$_2$ `intimately' mixed with H$_2$O ice and methanol ice (CH$_3$OH) detected a broad band centered near 2.134 $\micron$ \citep{bernstein2005near}, which is a factor of $\sim$10$^4$ weaker in pure CO$_2$ ice isolated in Ar \citep[]{sandford1991laboratory} or N$_2$ \citep[]{quirico1997near} matrices and is absent from pure H$_2$O ice. This band could result from the 2$\nu$$_3$ `forbidden' first overtone of the asymmetric stretching mode of CO$_2$ ice, centered near 2.134 $\micron$ \citep{bernstein2005near}, and could be  a useful indicator of the presence of well mixed CO$_2$ ice grains. 

 Two spectra collected at similar mid-observation, sub-observer longitudes and latitudes on Ariel's trailing hemisphere exhibit subtle absorption features centered near 2.129 and 2.130 $\micron$ (Figure 5), hereafter referred to as the `2.13-$\micron$' feature. These two spectra were collected by different teams using different telescopes (IRTF and ARC 3.5m), lending additional confidence to the detection of this 2.13-$\micron$ feature. The 2.13-$\micron$ bands we detected span 2.123 to 2.137 $\micron$ (0.014 $\micron$ wide) along the long wavelength shoulder of the 2.02-$\micron$ H$_2$O ice band. The CO$_2$ 2$\nu$$_3$ band presented in Fig. 3 of  \citet{bernstein2005near} is centered between $\sim$2.133 to 2.138 and is $\sim$0.016 to 0.041 $\micron$ wide, with the band becoming narrower and shifting to shorter wavelengths as the amount of CO$_2$ ice relative to H$_2$O ice is decreased. The closest match to the 2.13-$\micron$ band we detected on Ariel is a laboratory spectrum of a H$_2$O:CO$_2$:CH$_3$OH substrate (100:2.5:1 abundance ratio), displaying a band centered near 2.133 $\micron$ that is $\sim$0.016 $\micron$ wide \citet{bernstein2005near}. Using this laboratory substrate as a guide, if CO$_2$ is a well mixed constituent in Ariel's regolith, then it is possibly only a minor component compared to dirty H$_2$O ice. 
 
 If CO$_2$ ice is well mixed with dirty H$_2$O ice in Ariel's regolith, then these CO$_2$ ice grains could be a tracer of a radiolytic production cycle. In this scenario, CO$_2$ molecules, and other oxidized carbon compounds, would be produced by charged particle bombardment of Ariel's regolith. The charged particles would break apart H$_2$O molecules, forming OH radicals that interact with nearby C-rich grains to form CO$_2$ molecules:
\begin{center}OH + C(s) $\rightarrow$ CO + H\\
	OH + CO $\rightarrow$ CO$_2$ + H,\end{center}

\hspace{-0.3cm}where H efficiently diffuses out of Ariel's regolith and is preferentially removed via Jean's escape. Laboratory experiments \citep[e.g.,][]{raut2012radiation} have demonstrated that OH is an efficient catalyst for the production of CO$_2$ molecules in substrates composed of H$_2$O ice and C-rich material that are irradiated at cryogenic temperatures ($<$ 100 K) relevant to the Uranian moons. Supporting the possible presence of OH on the Uranian moons, a 280-nm absorption band, attributed to trapped OH, has been detected at high southern latitudes on Ariel ($\sim$45$\degree$S) using the Faint Object Spectrograph on the Hubble Space Telescope (HST) \citep[]{roush1998ultraviolet}. 

Other laboratory experiments have demonstrated that irradiation of hydrogenated carbon grains (i.e., carbon grains encased in H$_2$O ice) can also lead to radiolytic generation of CO$_2$ molecules \citep[e.g.,][]{mennella2004formation,mennella2006synthesis}. Such an `intramixture' of H$_2$O ice with embedded carbon grains could represent another useful analog for understanding the spectral signature of radiolytic chemistry in Ariel's regolith. 
	
	\begin{figure}[h!]
		\includegraphics[scale=0.9]{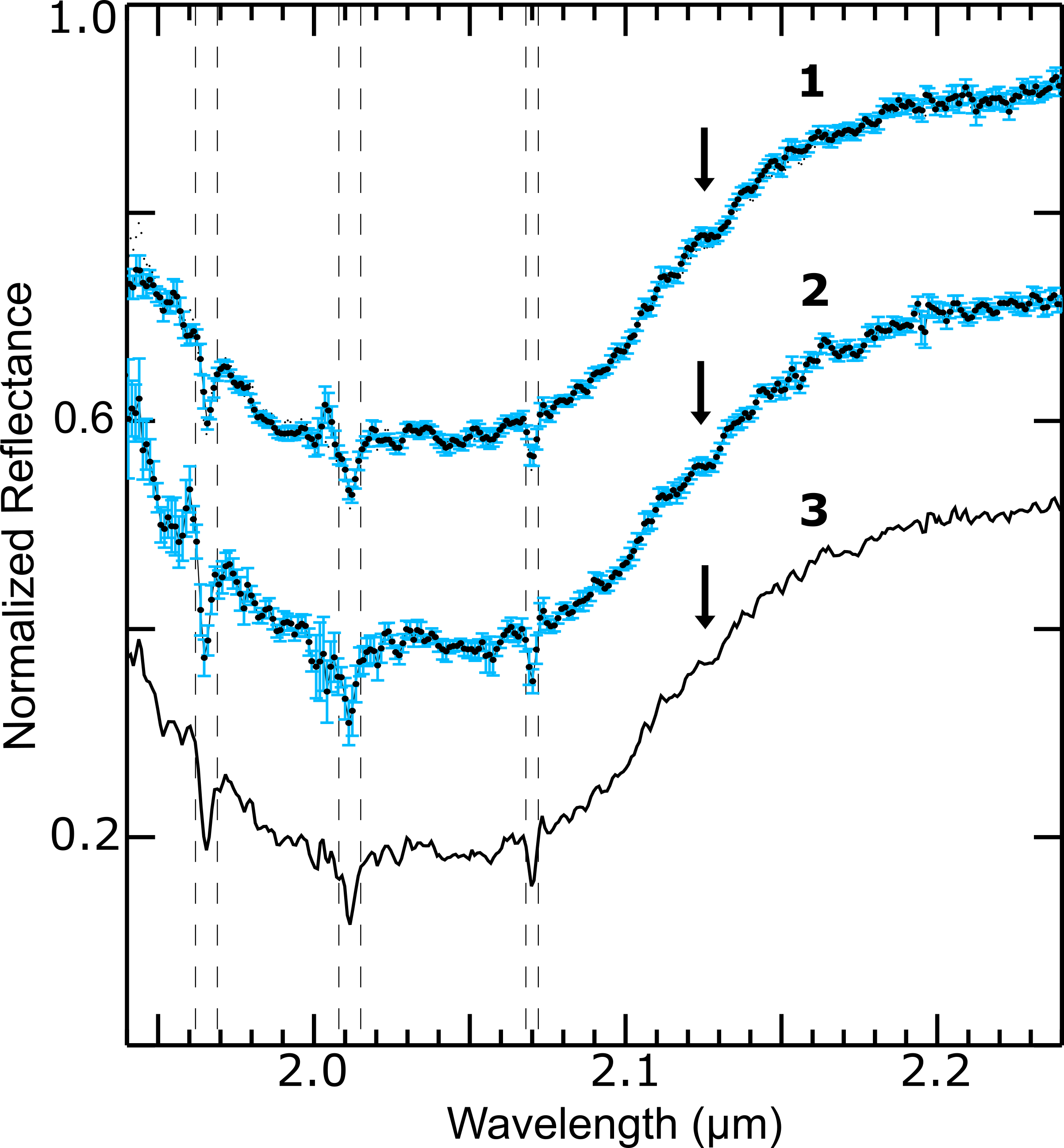}
		\caption{\textit{NIR spectra and 1$\sigma$ uncertainties (blue error bars) displaying evidence for a subtle absorption band centered near 2.13 $\micron$ (black arrows), located on the long wavelength shoulder of the 2.02-$\micron$ H$_2$O ice band. These two spectra were collected at mid-observation longitudes and latitudes of (1) 231.8$\degree$ and 51.1$\degree$N (IRTF/SpeX) and (2) 235.9$\degree$ and 47.3$\degree$N (ARC 3.5m/TripleSpec). The 2.13-$\micron$ band spans 2.123 to 2.137 $\micron$ in these two spectra with continuum-divided band depths of: (1) 1.8 $\pm$ 0.5$\%$ and (2) 1.7 $\pm$ 0.6$\%$. The bottom spectrum (3) is an average of these two spectra, highlighting the 2.13-$\micron$ band. The positions of CO$_2$ bands 1, 2, and 3 are delineated as well (black dashed lines). All spectra are normalized to 1 using the mean reflectance between 1.74 and 1.77 $\micron$ and offset vertically for clarity. These data were lightly smoothed using a 7 pixel wide boxcar function.}}\vspace{0.2 cm}
	\end{figure} 

In summary, a subtle 2.35-$\micron$ band we detected in near-infrared reflectance spectra collected over Ariel's trailing hemisphere hints at the presence of CO, which if present, is probably a transient species. The presence of a 2.13-$\micron$ band in two spectra collected at similar sub-observer latitudes and longitudes suggests that CO$_2$ molecules might be well mixed with H$_2$O ice in some high latitude locations. Such a mixture could be a tracer of radiolytic production sites where CO$_2$ molecules are being generated by irradiation of H$_2$O and C-rich material in Ariel's regolith. Particulate mixtures of CO$_2$ ice, H$_2$O ice, and C-rich material, or alternatively, intramixtures where CO$_2$ ice is formed at the boundary between C grains embedded in H$_2$O ice, could both represent suitable analogs for Ariel's regolith. Our results therefore support the presence of a sustained CO$_2$ `life cycle' on Ariel, with radiolytic production of CO$_2$ molecules and migration to low latitude cold traps, as predicted by thermodynamical models  \citep[]{grundy2006distributions,sori2017wunda}. 

It is also possible that CO$_2$ ice on Ariel originates in its interior and has been exposed by geologic processes, perhaps concentrated in deposits that are also rich in carbonates and/or NH$_3$-bearing species that could be contributing to Ariel's 2.2-$\micron$ band \citep{cartwright2020evidence}. However, the 2.2-$\micron$ band has been identified across Ariel's surface, possibly consistent with concentration in local geologic landforms or regional provinces, whereas CO$_2$ ice is present primarily on Ariel's trailing hemisphere and is essentially absent from its leading hemisphere. It is therefore difficult to explain the strong leading/trailing hemispherical asymmetries in the distribution of CO$_2$ ice on Ariel and the other Uranian moons if this constituent primarily originates from geologic processes.

	
	\subsection{Compositional stratification of Ariel's regolith?}
	\label{stratification}
	Analysis of synthetic spectra generated by prior spectral modeling efforts suggests that Ariel's surface could be compositionally stratified, with a thin layer of small H$_2$O ice grains capping concentrated deposits of CO$_2$ ice grains underneath \citep{cartwright2015distribution,cartwright2018red,cartwright2020probing}. These spectral modeling conclusions were based on the premise that near-infrared photons at different wavelengths travel different distances into planetary surfaces, depending on the composition of the grains within the regolith \citep[e.g.,][]{clark1984reflectance}. Comparison of a planetary body's spectral properties in different wavelength regions might therefore yield insight into the vertical structure of its regolith. One technique that can be used to gain such insight is the mean optical path length (MOPL), which provides an estimate for the average travel distance of photons into a particulate-dominated regolith before they are absorbed, accounting for inter-grain boundary scattering (for calculation details, see \citealt{clark1984reflectance}). 
	
Utilizing prior estimates of the MOPL for the Uranian moons' regoliths \citep{cartwright2018red}, we find that photons penetrate to depths of $\sim$0.1 mm over the wavelength range of CO$_2$ bands 1, 2, and 3 (1.9 -- 2.1 $\micron$). In contrast, photons penetrate to depths of $\sim$0.2 mm  over the wavelength range of the 1.543-$\micron$ and 1.578-$\micron$ CO$_2$ bands and $\sim$0.4 mm over the wavelength range of the 1.609-$\micron$ CO$_2$ band. The CO$_2$ ice grains included in synthetic spectra 1 -- 7, which fit the CO$_2$ bands between 1.5 to 1.7 $\micron$, are larger than the CO$_2$ grains included in synthetic spectra 8 -- 10, which fit CO$_2$ bands 1, 2, and 3 between 1.9 and 2.1 $\micron$. The larger CO$_2$ grains included in synthetic spectra 1 -- 7 could be simulating the factor of 2 to 4 longer path lengths that photons achieve through Ariel's regolith between 1.5 and 1.7 $\micron$. Relatively stronger absorption by H$_2$O ice between 1.9 and 2.1 $\micron$ reduces photon path lengths, thereby possibly explaining the smaller CO$_2$ grains included in the synthetic spectra that fit CO$_2$ bands 1, 2, and 3. Thus, our spectral modeling results are consistent with prior studies that suggest the Uranian moons' regoliths are compositionally stratified, with larger grains of CO$_2$ ice retained at greater depths \citep{cartwright2015distribution,cartwright2018red,cartwright2020probing}. 

The spectral modeling results presented here might instead hint at a different regolith structure, where CO$_2$ ice is formed beneath a thin H$_2$O ice cap, at the boundary between C grains embedded into a dominantly H$_2$O ice matrix (i.e., intramixture). Albeit, visible polarimetric observations of the Uranian moons are consistent with crumbly regoliths that include sub-micron sized grains \citep{afanasiev2014polarimetry}, supporting the interpretation that particulate mixtures are good analogs for Ariel's regolith. Determining whether particulate mixtures or intramixtures are  better analogs for the structure of Ariel's regolith is beyond the scope of this work and should be considered by future studies. As described in Section 3.4, minor variations in grain size and fractional coverage in each of synthetic spectra 1 -- 7 do not cause significant changes in their goodness-of-fit to the CO$_2$ bands between 1.5 and 1.7 $\micron$ (Table 5). Consequently, these synthetic spectra may not be entirely sensitive to the factor of $\sim$2 change in photon penetration depth between the 1.543-$\micron$ and 1.578-$\micron$ bands and the 1.609-$\micron$ band described above. 
	
	\section{Conclusions and Future Work} 
	
	We measured the distribution and spectral signature of CO$_2$ ice as a function of sub-observer latitude and longitude to investigate whether a radiolytic production cycle, and associated transport of CO$_2$ molecules to low latitude cold traps, is occurring on Ariel. We measured the integrated areas of three CO$_2$ ice bands between 1.9 and 2.1 $\micron$ in 31 reflectance spectra collected over a wide range of sub-observer longitudes and latitudes (30$\degree$S -- 51$\degree$N). Our results and analyses indicate that CO$_2$ ice band areas are decreasing with increasing sub-observer latitude on Ariel's trailing hemisphere but do not vary on its leading hemisphere. These results suggest that the amount of CO$_2$ ice on Ariel's surface is higher at low latitudes, as predicted by thermodynamical models. Furthermore, two spectra collected at high northern sub-observer latitudes (47$\degree$N and 51$\degree$N) display subtle 2.13-$\micron$ bands, hinting at the presence of CO$_2$ ice grains well mixed with H$_2$O ice and carbonaceous material, possibly representing a spectral tracer of radiolytic production sites for CO$_2$ molecules in Ariel's regolith. Similarly, a subtle 2.35-$\micron$ band detected in low sub-observer latitude spectra of Ariel hints at the presence of CO, which could be a transient product generated by irradiation and fragmentation of CO$_2$ molecules in low latitude cold traps. The work presented here is therefore consistent with radiolytic production of CO$_2$ molecules and subsequent migration and cold trapping of this constituent in concentrated deposits of CO$_2$ ice at low latitudes on Ariel. 

	Continued observations of Ariel over the next decade as the Uranian system approaches northern summer solstice, when the sub-solar latitude will reach 82$\degree$N, will provide key measurements of the changing distribution and spectral signature of CO$_2$ ice. These future measurements could be compared to the results presented here to further test whether CO$_2$ ice is a radiolytic product that migrates to low latitude cold traps. Furthermore, follow up observations are needed to confirm the presence of the 2.13-$\micron$ band reported here, and determine whether this feature is present in other locations. Similar to Ariel, CO$_2$ ice has been detected in reflectance spectra collected over low sub-observer latitudes on the trailing hemispheres of Umbriel, Titania, and Oberon \citep{grundy2003discovery,grundy2006distributions,cartwright2015distribution}, but the nature of CO$_2$ ice at high sub-observer latitudes on these moons has yet to be investigated. If CO$_2$ ice originates as part of a radiolytic production cycle on Umbriel, Titania, and Oberon, then CO$_2$ ice band areas should decrease in spectra collected at higher sub-observer latitudes on these moons, similar to the observed trends on Ariel. Furthermore, determining whether subtle 2.13-$\micron$ bands are present in spectra collected over high sub-observer latitudes of these three moons could provide insight into whether this feature is a tracer of radiolytic production sites in the Uranian satellites' regoliths. Follow up near-UV observations are also needed to confirm whether a 280-nm band attributed to trapped OH, a key catalyst for the radiolytic production of CO$_2$ molecules, is present at high northern latitudes on Ariel and the other Uranian moons. 
			
	To fully characterize the life cycle of CO$_2$ ice on the classical Uranian satellites, a spacecraft equipped with a visible wavelength camera and near-infrared mapping spectrometer, making multiple close flybys of these moons, is needed to reveal their previously unimaged northern hemispheres, measure the localized distribution and spectral signature of CO$_2$ ice, and characterize any possible associations between CO$_2$ ice and geologic landforms like craters and chasmata that might serve as cold traps \citep[e.g.,][]{cartwright2021sciencecase,leonard2021umami}. Similarly, a Uranus orbiter equipped with a plasma spectrometer and an energetic particle detector could measure Uranus' magnetosphere proximal to Ariel and the other moons, thereby providing critical context on moon-magnetosphere interactions that could be driving CO$_2$ production and volatile cycling on these moons \citep[e.g.,][]{kollmann2020magnetospheric}. New measurements made by a thermal camera onboard a Uranus orbiter are also needed to measure local and regional variations in the surface temperatures of these moons, which would provide important information on the location and longevity of CO$_2$ ice cold traps and the source regions where CO$_2$ molecules are being generated by radiolysis or exposed by geologic processes.
		
	\section{Acknowledgments} 
	This project was funded by the JPL Research and Technology Development Fund. Portions of this work were carried out at the Jet Propulsion Laboratory, California Institute of Technology, under contract to the National Aeronautics and Space Administration. Many of the observations reported here were made from the summit of Maunakea, and we thank the people of Hawaii for the opportunity to observe from this special mountain. Feedback and edits provided by two anonymous reviewers were incorporated into this paper.

	\bibliography{references}{}

\begin{thebibliography}{}
\expandafter\ifx\csname natexlab\endcsname\relax\def\natexlab#1{#1}\fi
\providecommand{\url}[1]{\href{#1}{#1}}
\providecommand{\dodoi}[1]{doi:~\href{http://doi.org/#1}{\nolinkurl{#1}}}
\providecommand{\doeprint}[1]{\href{http://ascl.net/#1}{\nolinkurl{http://ascl.net/#1}}}
\providecommand{\doarXiv}[1]{\href{https://arxiv.org/abs/#1}{\nolinkurl{https://arxiv.org/abs/#1}}}

\bibitem[{Afanasiev {et~al.}(2014)Afanasiev, Rosenbush, \&
  Kiselev}]{afanasiev2014polarimetry}
Afanasiev, V., Rosenbush, V., \& Kiselev, N. 2014, Astrophysical Bulletin, 69,
  211

\bibitem[{Andrae {et~al.}(2010)Andrae, Schulze-Hartung, \&
  Melchior}]{andrae2010and}
Andrae, R., Schulze-Hartung, T., \& Melchior, P. 2010, arXiv preprint
  arXiv:1012.3754

\bibitem[{Beddingfield {et~al.}(2015)Beddingfield, Burr, \&
  Emery}]{beddingfield2015fault}
Beddingfield, C., Burr, D., \& Emery, J. 2015, Icarus, 247, 35

\bibitem[{Beddingfield \& Cartwright(2020)}]{beddingfield2020hidden}
Beddingfield, C.~B., \& Cartwright, R.~J. 2020, Icarus, 113687

\bibitem[{Beddingfield \& Cartwright(2021)}]{beddingfield2021Arielcryo}
---. 2021, Icarus, 114583

\bibitem[{Bernstein {et~al.}(2005)Bernstein, Cruikshank, \&
  Sandford}]{bernstein2005near}
Bernstein, M.~P., Cruikshank, D.~P., \& Sandford, S.~A. 2005, Icarus, 179, 527

\bibitem[{Bertrand \& Forget(2016)}]{bertrand2016observed}
Bertrand, T., \& Forget, F. 2016, Nature, 540, 86

\bibitem[{Bevington \& Robinson(1969)}]{bevington1969data}
Bevington, P., \& Robinson, D.~K. 1969, Data Analysis and Error Reduction for
  the Physical Sciences,  McGraw-Hill, New York

\bibitem[{Bohren \& Huffman(1983)}]{bohren1983light}
Bohren, C., \& Huffman, D. 1983, Wiley, New York

\bibitem[{Brown \& Clark(1984)}]{brown1984surface}
Brown, R.~H., \& Clark, R.~N. 1984, Icarus, 58, 288

\bibitem[{Brown \& Cruikshank(1983)}]{brown1983uranian}
Brown, R.~H., \& Cruikshank, D.~P. 1983, Icarus, 55, 83

\bibitem[{Buratti \& Mosher(1991)}]{buratti1991comparative}
Buratti, B.~J., \& Mosher, J.~A. 1991, Icarus, 90, 1

\bibitem[{Cartwright {et~al.}(2021)Cartwright, Beddingfield, Nordheim, Elder,
  Castillo-Rogez, Bramson, Sori, Buratti, Neveu, Pappalardo, Cohen, Leonard,
  Ermakov, Roser, Hofstadter, Showalter, Grundy, \&
  Turtle}]{cartwright2021sciencecase}
Cartwright, R., Beddingfield, C., Nordheim, T., {et~al.} 2021, Planetary
  Science Journal, 2, 120

\bibitem[{Cartwright {et~al.}(2020{\natexlab{a}})Cartwright, Emery, Grundy,
  Cruikshank, Beddingfield, \& Pinilla-Alonso}]{cartwright2020probing}
Cartwright, R.~J., Emery, J.~P., Grundy, W.~M., {et~al.} 2020{\natexlab{a}},
  Icarus, 338, 113513

\bibitem[{Cartwright {et~al.}(2018)Cartwright, Emery, Pinilla-Alonso, Lucas,
  Rivkin, \& Trilling}]{cartwright2018red}
Cartwright, R.~J., Emery, J.~P., Pinilla-Alonso, N., {et~al.} 2018, Icarus,
  314, 210

\bibitem[{Cartwright {et~al.}(2015)Cartwright, Emery, Rivkin, Trilling, \&
  Pinilla-Alonso}]{cartwright2015distribution}
Cartwright, R.~J., Emery, J.~P., Rivkin, A.~S., Trilling, D.~E., \&
  Pinilla-Alonso, N. 2015, Icarus, 257, 428

\bibitem[{Cartwright {et~al.}(2020{\natexlab{b}})Cartwright, Beddingfield,
  Nordheim, Roser, Grundy, Hand, Emery, Cruikshank, \&
  Scipioni}]{cartwright2020evidence}
Cartwright, R.~J., Beddingfield, C.~B., Nordheim, T.~A., {et~al.}
  2020{\natexlab{b}}, The Astrophysical Journal Letters, 898, L22

\bibitem[{Clark \& Lucey(1984)}]{clark1984spectral}
Clark, R.~N., \& Lucey, P.~G. 1984, Journal of Geophysical Research: Solid
  Earth, 89, 6341

\bibitem[{Clark \& Roush(1984)}]{clark1984reflectance}
Clark, R.~N., \& Roush, T.~L. 1984, Journal of Geophysical Research: Solid
  Earth, 89, 6329

\bibitem[{Combe {et~al.}(2019)Combe, McCord, Matson, Johnson, Davies, Scipioni,
  \& Tosi}]{combe2019nature}
Combe, J.-P., McCord, T.~B., Matson, D.~L., {et~al.} 2019, Icarus, 317, 491

\bibitem[{Cook {et~al.}(2018)Cook, Dalle~Ore, Protopapa, Binzel, Cartwright,
  Cruikshank, Earle, Grundy, Ennico, Howett, {et~al.}}]{cook2018composition}
Cook, J.~C., Dalle~Ore, C.~M., Protopapa, S., {et~al.} 2018, Icarus, 315, 30

\bibitem[{Cruikshank {et~al.}(1977)Cruikshank, Pilcher, \&
  Morrison}]{cruikshank1977identification}
Cruikshank, D., Pilcher, C.~B., \& Morrison, D. 1977, The Astrophysical
  Journal, 217, 1006

\bibitem[{Cruikshank(1980)}]{cruikshank1980near}
Cruikshank, D.~P. 1980, Icarus, 41, 246

\bibitem[{Cruikshank \& Brown(1981)}]{cruikshank1981uranian}
Cruikshank, D.~P., \& Brown, R.~H. 1981, Icarus, 45, 607

\bibitem[{Cushing {et~al.}(2004)Cushing, Vacca, \&
  Rayner}]{cushing2004spextool}
Cushing, M.~C., Vacca, W.~D., \& Rayner, J.~T. 2004, Publications of the
  Astronomical Society of the Pacific, 116, 362

\bibitem[{DeColibus {et~al.}(2020)DeColibus, Chanover, \&
  Cartwright}]{decolibus2020investigating}
DeColibus, D., Chanover, N., \& Cartwright, R. 2020, AAS/Division for Planetary
  Sciences Meeting, 52, 215

\bibitem[{Earle {et~al.}(2017)Earle, Binzel, Young, Stern, Ennico, Grundy,
  Olkin, Weaver, {et~al.}}]{earle2017long}
Earle, A.~M., Binzel, R.~P., Young, L.~A., {et~al.} 2017, Icarus, 287, 37

\bibitem[{Emery {et~al.}(2006)Emery, Cruikshank, \&
  Van~Cleve}]{emery2006thermal}
Emery, J., Cruikshank, D., \& Van~Cleve, J. 2006, Icarus, 182, 496

\bibitem[{Farmer {et~al.}(1976)Farmer, Davies, \& LaPorte}]{farmer1976mars}
Farmer, C.~B., Davies, D.~W., \& LaPorte, D.~D. 1976, Science, 194, 1339

\bibitem[{Gerakines {et~al.}(2001)Gerakines, Moore, \&
  Hudson}]{gerakines2001energetic}
Gerakines, P., Moore, M., \& Hudson, R. 2001, Journal of Geophysical Research:
  Planets, 106, 33381

\bibitem[{Gerakines {et~al.}(2005)Gerakines, Bray, Davis, \&
  Richey}]{gerakines2005strengths}
Gerakines, P.~A., Bray, J., Davis, A., \& Richey, C. 2005, The Astrophysical
  Journal, 620, 1140

\bibitem[{Grundy {et~al.}(2006)Grundy, Young, Spencer, Johnson, Young, \&
  Buie}]{grundy2006distributions}
Grundy, W., Young, L., Spencer, J., {et~al.} 2006, Icarus, 184, 543

\bibitem[{Grundy {et~al.}(2003)Grundy, Young, \& Young}]{grundy2003discovery}
Grundy, W., Young, L., \& Young, E. 2003, Icarus, 162, 222

\bibitem[{Hamilton {et~al.}(2016)Hamilton, Stern, Moore, \&
  Young}]{hamilton2016rapid}
Hamilton, D.~P., Stern, S., Moore, J., \& Young, L. 2016, Nature, 540, 97

\bibitem[{Hanel {et~al.}(1986)Hanel, Conrath, Flasar, Kunde, Maguire, Pearl,
  Pirraglia, Samuelson, Cruikshank, Gautier, {et~al.}}]{hanel1986infrared}
Hanel, R., Conrath, B., Flasar, F., {et~al.} 1986, Science, 233, 70

\bibitem[{Hansen(1997)}]{hansen1997spectral}
Hansen, G.~B. 1997, Advances in Space Research, 20, 1613

\bibitem[{Hansen(2005)}]{hansen2005ultraviolet}
---. 2005, Journal of Geophysical Research: Planets, 110

\bibitem[{Hapke(2012)}]{hapke2012theory}
Hapke, B. 2012, Theory of reflectance and emittance spectroscopy (Cambridge
  university press)

\bibitem[{Hendrix {et~al.}(2019)Hendrix, Hurford, Barge, Bland, Bowman,
  Brinckerhoff, Buratti, Cable, Castillo-Rogez, Collins,
  {et~al.}}]{hendrix2019nasa}
Hendrix, A.~R., Hurford, T.~A., Barge, L.~M., {et~al.} 2019, Astrobiology, 19,
  1

\bibitem[{Holler {et~al.}(2016)Holler, Young, Grundy, \&
  Olkin}]{holler2016surface}
Holler, B., Young, L., Grundy, W., \& Olkin, C. 2016, Icarus, 267, 255

\bibitem[{Karkoschka(2001)}]{karkoschka2001comprehensive}
Karkoschka, E. 2001, Icarus, 151, 51

\bibitem[{Kieffer(1979)}]{kieffer1979mars}
Kieffer, H.~H. 1979, Journal of Geophysical Research: Solid Earth, 84, 8263

\bibitem[{Kieffer {et~al.}(2000)Kieffer, Titus, Mullins, \&
  Christensen}]{kieffer2000mars}
Kieffer, H.~H., Titus, T.~N., Mullins, K.~F., \& Christensen, P.~R. 2000,
  Journal of Geophysical Research: Planets, 105, 9653

\bibitem[{Kollmann {et~al.}(2020)Kollmann, Cohen, Allen, Clark, Roussos, Vines,
  Dietrich, Wicht, de~Pater, Runyon, {et~al.}}]{kollmann2020magnetospheric}
Kollmann, P., Cohen, I., Allen, R., {et~al.} 2020, Space Science Reviews, 216,
  1

\bibitem[{Leonard {et~al.}(2021)Leonard, Elder, Nordheim, Cartwright, Patthoff,
  Beddingfield, Cochrane, Brooks, Tiscareno, Strange,
  {et~al.}}]{leonard2021umami}
Leonard, E.~J., Elder, C., Nordheim, T.~A., {et~al.} 2021, The Planetary
  Science Journal, 2, 174

\bibitem[{Mastrapa {et~al.}(2008)Mastrapa, Bernstein, Sandford, Roush,
  Cruikshank, \& Dalle~Ore}]{mastrapa2008optical}
Mastrapa, R., Bernstein, M., Sandford, S., {et~al.} 2008, Icarus, 197, 307

\bibitem[{Mennella {et~al.}(2006)Mennella, Baratta, Palumbo, \&
  Bergin}]{mennella2006synthesis}
Mennella, V., Baratta, G., Palumbo, M., \& Bergin, E. 2006, The Astrophysical
  Journal, 643, 923

\bibitem[{Mennella {et~al.}(2004)Mennella, Palumbo, \&
  Baratta}]{mennella2004formation}
Mennella, V., Palumbo, M., \& Baratta, G. 2004, The Astrophysical Journal, 615,
  1073

\bibitem[{Paranicas {et~al.}(1996)Paranicas, Cheng, \&
  Mauk}]{paranicas1996charged}
Paranicas, C., Cheng, A.~F., \& Mauk, B.~H. 1996, Journal of Geophysical
  Research: Space Physics, 101, 10681

\bibitem[{Quirico \& Schmitt(1997)}]{quirico1997near}
Quirico, E., \& Schmitt, B. 1997, Icarus, 127, 354

\bibitem[{Raut {et~al.}(2012)Raut, Fulvio, Loeffler, \&
  Baragiola}]{raut2012radiation}
Raut, U., Fulvio, D., Loeffler, M., \& Baragiola, R. 2012, The Astrophysical
  Journal, 752, 159

\bibitem[{Rayner {et~al.}(2003)Rayner, Toomey, Onaka, Denault, Stahlberger,
  Vacca, Cushing, \& Wang}]{rayner2003spex}
Rayner, J., Toomey, D., Onaka, P., {et~al.} 2003, Publications of the
  Astronomical Society of the Pacific, 115, 362

\bibitem[{Rouleau \& Martin(1991)}]{rouleau1991shape}
Rouleau, F., \& Martin, P. 1991, The Astrophysical Journal, 377, 526

\bibitem[{Roush {et~al.}(1998)Roush, Noll, Cruikshank, \&
  Pendleton}]{roush1998ultraviolet}
Roush, T.~L., Noll, K.~S., Cruikshank, D.~P., \& Pendleton, Y. 1998, in Lunar
  and Planetary Science Conference No. 1636, 1636

\bibitem[{Sandford {et~al.}(1991)Sandford, Salama, Allamandola, Trafton,
  Lester, \& Ramseyer}]{sandford1991laboratory}
Sandford, S.~A., Salama, F., Allamandola, L.~J., {et~al.} 1991, Icarus, 91, 125

\bibitem[{Schenk \& Moore(2020)}]{schenk2020topography}
Schenk, P.~M., \& Moore, J.~M. 2020, Philosophical Transactions of the Royal
  Society A, 378, 20200102

\bibitem[{Smith {et~al.}(1986)Smith, Soderblom, Beebe, Bliss, Boyce, Brahic,
  Briggs, Brown, Collins, Cook, {et~al.}}]{smith1986voyager}
Smith, B.~A., Soderblom, L., Beebe, R., {et~al.} 1986, Science, 233, 43

\bibitem[{Soifer {et~al.}(1981)Soifer, Neugebauer, \&
  Matthews}]{soifer1981near}
Soifer, B., Neugebauer, G., \& Matthews, K. 1981, Icarus, 45, 612

\bibitem[{Sori {et~al.}(2017)Sori, Bapst, Bramson, Byrne, \&
  Landis}]{sori2017wunda}
Sori, M.~M., Bapst, J., Bramson, A.~M., Byrne, S., \& Landis, M.~E. 2017,
  Icarus, 290, 1

\bibitem[{Spencer \& Calvin(2002)}]{spencer2002condensed}
Spencer, J.~R., \& Calvin, W.~M. 2002, The Astronomical Journal, 124, 3400

\bibitem[{Spencer {et~al.}(1995)Spencer, Calvin, \& Person}]{spencer1995charge}
Spencer, J.~R., Calvin, W.~M., \& Person, M.~J. 1995, Journal of Geophysical
  Research: Planets, 100, 19049

\bibitem[{Waite {et~al.}(2006)Waite, Combi, Ip, Cravens, McNutt, Kasprzak,
  Yelle, Luhmann, Niemann, Gell, {et~al.}}]{waite2006cassini}
Waite, J.~H., Combi, M.~R., Ip, W.-H., {et~al.} 2006, science, 311, 1419

\bibitem[{Wilson {et~al.}(2004)Wilson, Henderson, Herter, Matthews, Skrutskie,
  Adams, Moon, Smith, Gautier, Ressler, {et~al.}}]{wilson2004mass}
Wilson, J.~C., Henderson, C.~P., Herter, T.~L., {et~al.} 2004, in Ground-based
  Instrumentation for Astronomy, Vol. 5492, International Society for Optics
  and Photonics, 1295--1305

\bibitem[{Zheng \& Kaiser(2007)}]{zheng2007formation}
Zheng, W., \& Kaiser, R.~I. 2007, Chemical Physics Letters, 450, 55

\end{thebibliography}
	\bibliographystyle{aasjournal}
	
	
	
	\renewcommand{\thesubsection}{A\arabic{subsection}}
	\setcounter{subsection}{0}
	
	\renewcommand{\thefigure}{A\arabic{figure}}
	\setcounter{figure}{0}
	
	\renewcommand{\thetable}{A\arabic{table}}
	\setcounter{table}{0}
	
	\section*{Appendix}
	
	\subsection{Methods: Band parameter measurements}
	In this appendix, we show the wavelength ranges of CO$_2$ ice bands 1, 2, and 3 and their associated continua. 
	\begin{figure}[h!]
		\includegraphics[scale=0.7]{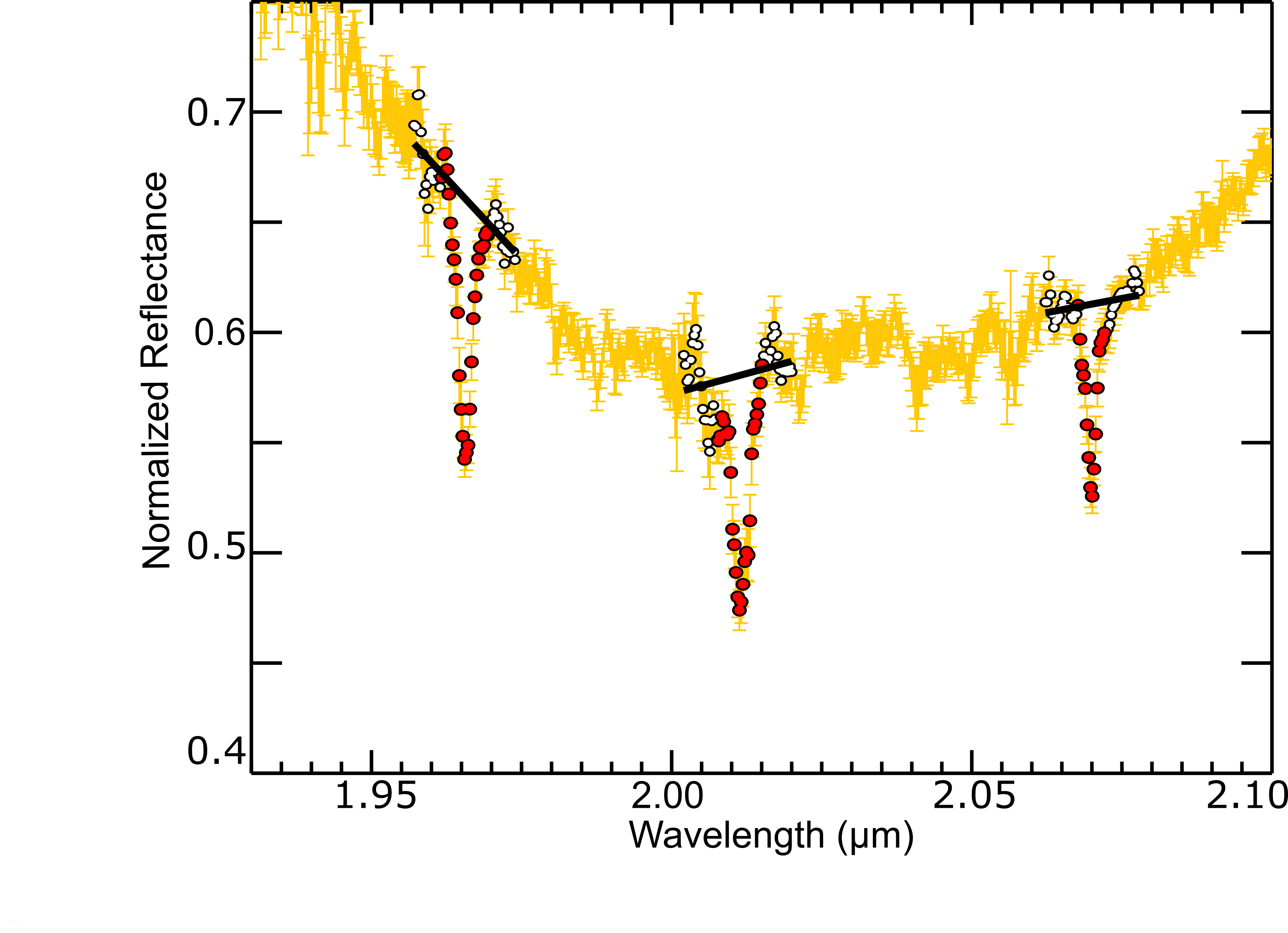}
		\caption{\textit{A graphical demonstration of the wavelength ranges of CO$_2$ ice bands 1, 2, and 3, and their associated continua, using the grand average low latitude spectrum of Ariel (1$\sigma$ uncertainties shown as gold error bars). The data points used to calculate the areas of CO$_2$ bands 1, 2, and 3 (spanning 1.962 -- 1.969 $\micron$, 2.008 -- 2.015 $\micron$, and 2.068 -- 2.072 $\micron$, respectively) are shown as red  data points. The data points utilized for the continua of these three CO$_2$ bands are shown as white data points that are connected by black lines.}}\vspace{0.1 cm}
	\end{figure} 
	
	\clearpage
	
	\subsection{Results: IRTF/SpeX and ARC 3.5 m/TripleSpec spectra}
	Here we show ten near-infrared reflectance spectra of the Uranian satellite Ariel at their native spectral resolutions.\vspace{-0.4 cm} 
	\begin{figure}[h!]
		\includegraphics[scale=0.64]{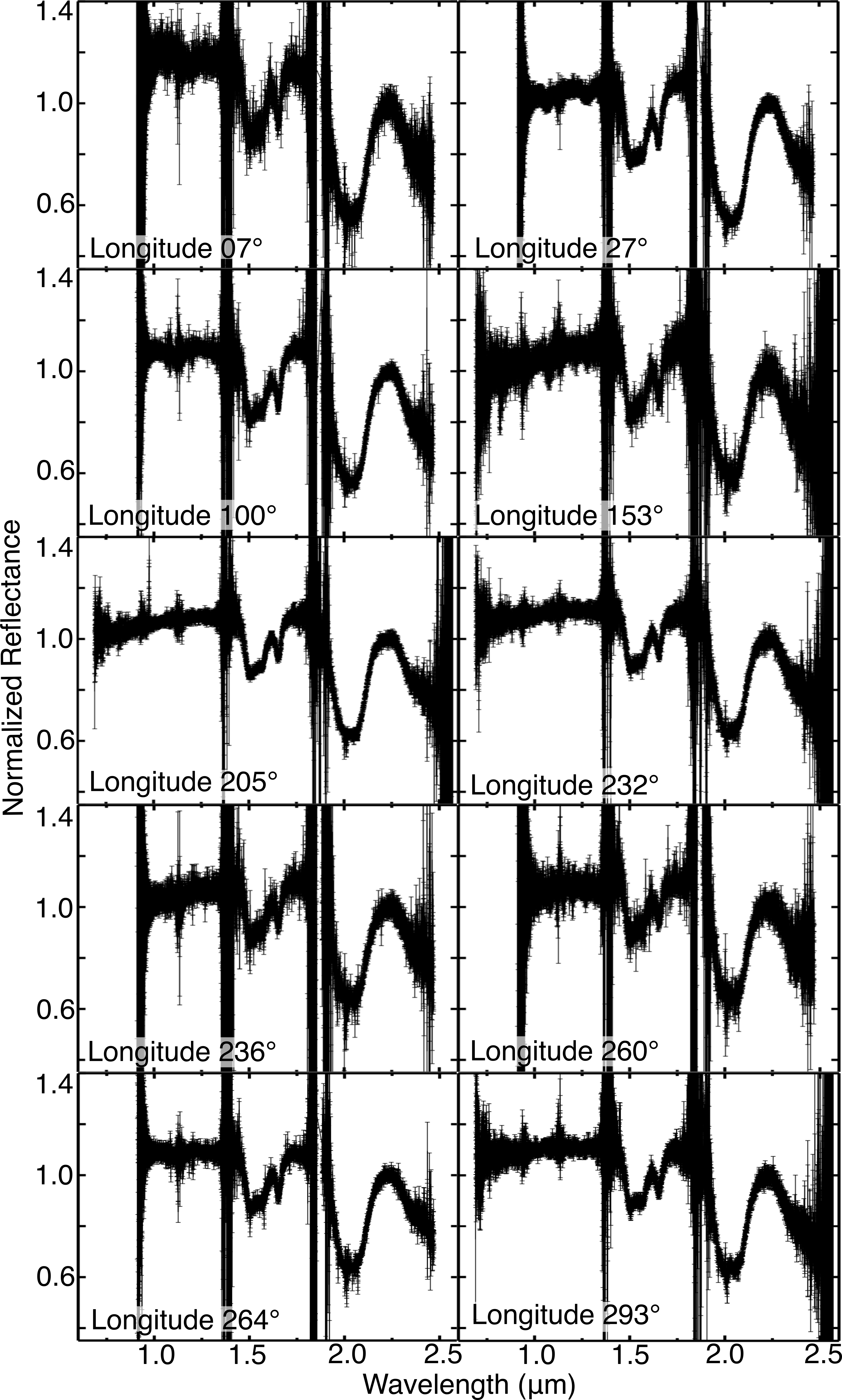}
		\caption{\textit{Four new IRTF/SpeX spectra (longitudes 153$\degree$, 205$\degree$, 232$\degree$, and 293$\degree$) and six new ARC 3.5m/TripleSpec spectra (longitudes 07$\degree$, 27$\degree$, 100$\degree$, 236$\degree$, 260$\degree$, and 264$\degree$) of Ariel and their 1$\sigma$ uncertainties, collected in 2019 and 2020. The mid-observation, sub-observer longitude for each spectrum is included in the bottom left-hand corner of each plot (see Table 1 for observation details). All spectra have been normalized to 1 between 2.24 to 2.25 $\micron$. The sensitivities of SpeX (SXD mode) and TripleSpec are lower at wavelengths $<$ 1.4 $\micron$ and $>$ 2.3 $\micron$, resulting in lower S/N compared to wavelengths between 1.4 to 2.3 $\micron$. Scattered light from Uranus contributes additional structure to some of these spectra at wavelengths $<$ 1.3 $\micron$.}}\vspace{1 cm}
	\end{figure} 
\end{document}